\documentclass[10pt,journal,compsoc]{IEEEtran}

\usepackage[T1]{fontenc}
\usepackage{amsmath}
\usepackage{graphicx}
\usepackage{array}
\usepackage{rotating}
\usepackage{xcolor}
\usepackage{soul,color}

\usepackage{subcaption}
 \usepackage{amsfonts}

\begin{document}
\title{Novel Wake-up Scheme for Energy-Efficient Low-Latency Mobile Devices in 5G Networks}
\author{Soheil~Rostami, Kari Heiska, Oleksandr Puchko, Kari Leppanen, and \\ Mikko Valkama,~\IEEEmembership{Senior Member,~IEEE}
\thanks{S. Rostami, K. Heiska, O. Puchko and K. Leppanen are with Huawei Technologies Oy (Finland)  Co. Ltd, Helsinki R\&D Center, Finland. 
}
\thanks{S. Rostami  and M. Valkama are with the
Department of Electrical Engineering, Tampere University, Finland. 
}
\thanks{Corresponding author: Mikko Valkama, mikko.valkama@tuni.fi}
\thanks{This work has received funding from the European Union's Horizon 2020 research and innovation program under the Marie Sk\l{}odowska-Curie grant agreement No. 675891 (SCAVENGE).}
\thanks{{Limited subset of early stage results were presented at IEEE
GLOBECOM, Dec 2018 \cite{Globecom}.} } 
}
\maketitle
\begin{abstract}
 Improved mobile device battery lifetime and latency minimization are critical requirements for enhancing the mobile broadband services and user experience. Long-term evolution (LTE) networks have adopted discontinuous reception (DRX)  as the baseline solution for prolonged battery lifetime. However, in every DRX cycle, the mobile device baseband  processing unit monitors and decodes the control signaling, and thus all instances without any actual data allocation leads to unnecessary energy consumption. This fact together with the long start-up and power-down times can prevent adopting frequent wake-up instants, which in turn leads to considerable latency.     In this work, a novel wake-up scheme is described and studied, to tackle the trade-off between latency and battery lifetime in future $5$G networks, seeking thus to facilitate an always-available experience, rather than always-on. Analytical and simulation-based  results show that the proposed scheme is a promising approach to control   the  user plane latency  and energy consumption, when the device is operating in the power saving mode. The aim of this article is to describe the overall wake-up system operating principle and the associated signaling methods, receiver processing solutions and essential implementation aspects. Additionally, the advantages compared to DRX-based systems are shown and demonstrated, through the analysis of the system energy-efficiency and latency characteristics, with special emphasis on future 5G-grade mobile devices.
\end{abstract}
\IEEEpeerreviewmaketitle

\section{Introduction}
  \IEEEPARstart{I}{n} future $5$G networks,  it is expected
  that a diverse set of services, e.g., augmented and virtual reality, cloud gaming, ultra-high-definition   video streaming and connected cars, with very aggressive  quality-of-service (QoS) requirements will become ubiquitous  \cite{Boccardi2014}.  For instance, in order to avoid virtual reality sickness as a result of using superimposed computer-generated content over live images viewed through smartphone cameras, acceptable {end-to-end} latency   is less than a few tens of milliseconds, and requires throughput in the order of  $100$ Mbps \cite{Alamouti}. On the other hand,  the slow technological progress  in battery capacities complicates adopting more and more complex and high-rate continuous processing in the devices, and therefore different energy saving features and mechanisms are of substantial importance \cite{Jaya}. To this end, multiple experimental investigations show that  the most energy consuming components of  smartphones can be attributed to the cellular connectivity subsystem and the  display \cite{Qualcomm2013,Carroll2010,Lauridsenthesis}.   Therefore, the design of an energy-efficient   cellular subsystem, utilizing the battery energy capacity in an efficient manner,  is extremely  paramount.  In this article, due to the asymmetric data traffic which is dominated by downlink \cite{ITU2015},   our main focus is on the device receive mode.

Since the last decade,  different mobile device energy saving features have been considered by different standardization and device implementation bodies. Particularly, the $3$rd generation partnership project ($3$GPP) has introduced  discontinuous reception (DRX) as an important energy saving mechanism for   long-term evolution (LTE) mobile networks \cite{erik}.   DRX   enables  the mobile device to reduce energy consumption by systematically turning some components of  the cellular subsystem  in sleep mode for certain periods of time. On the other hand, for low communication latencies, it is necessary to shorten the DRX cycle as much as possible, emulating an always-available experience. However, due to the existence of  relatively  long start-up and power-down periods,  as well as high energy consumption of the baseband processing unit (BBU) during the active time,
the DRX mechanism cannot alone boost the $5$G mobile devices battery lifetime sufficiently, especially when reduced latencies with good power-efficiency are pursued.

In general, the above limitations prevent designing and facilitating energy-efficient low-latency communications in future $5$G networks 
based on current DRX-based power saving mechanisms alone.  In recent years, a new type of approach, the so-called \textit{wake-up receiver (WRx)}  concept, has been introduced to improve   the lifetime of the nodes in   wireless sensor networks (WSNs) \cite{Mazloum2014}. In this article, the concept  of WRx and associated wake-up signaling are utilized and developed to prolong the mobile devices battery lifetime while still facilitating controlled latency. We refer to the proposed WRx-empowered cellular subsystem as the \emph{new modem (NM)} in the rest of this article. The extensive numerical results presented in the article show that the NM concept reduces the energy consumption of DRX-enabled cellular subsystem up to  {{$30$\%}},  depending on the traffic pattern and user requirements, while guaranteeing a predictable and consistent latency in the order of   $25$ milliseconds. Because of low complexity and  low power consumption of the WRx, together with the relatively low latency, the developed NM approach is not only applicable for   enhanced mobile broadband (eMBB) purposes, but   also for machine-centric use cases, such as  utility management, or potentially also for real-time command and control for remote medication and surgery,  in which   battery lifetime is critically important, while at the same time  demanding low latency communication \cite{Alamouti,Jaya}.

 The work reported in this article consists of developing a novel wake-up scheme for scheduled frame-based cellular radio systems, including the necessary wake-up signaling and detection processing, analyzing the achievable performance in terms of energy-efficiency and latency, as well as shortly addressing the needed processing integration into available LTE-optimized $16$ nm CMOS-based transceiver ICs. Specifically, building on our initial work in \cite{Globecom}, the major contributions of this {article} can be stated as follows 
 \begin{itemize}
 \item an efficient wake-up signaling structure is introduced, where different orthogonal sequences are utilized to distinguish among   different   WRxs;
\item   a  low-complexity WRx processing solution  is developed and analyzed  to detect  the corresponding sequence, and to acquire the associated time and frequency synchronization;

\item{average latency and power consumption of the proposed concept are quantified and characterized in an analytical manner using semi-markov process;}
  
\item   extensive link- and system-level simulations  are presented      to quantify the correctness and effectiveness of   the proposed scheme and analytical results, and to compare its performance with more ordinary DRX-based cellular subsystem.
\end{itemize}
 
  To the best of authors' knowledge, this work  is in the fore-front in developing such a wake-up based scheme for $5$G mobile devices, such that all relevant aspects covering the wake-up signal structure, detection processing, energy-consumption and latency analysis, implementation details and the achievable overall system performance are all systematically addressed. Overall, the proposed scheme and the article should be of interest to a broad readership including academic and industrial research communities, who are seeking feasible methods to reduce the energy consumption of  mobile devices.

The mathematical notations used in this work are as follows:  $I_{n}$ represents the $n \times n$ identity matrix, $(.)^{*}$ represent conjugate operation, $\hat{(.)}$ denotes a parameter estimate,   $(.)^{T}$ denotes the transpose operation, $\mathbb{E}[.]$ is the expectation   operation, $\Pr[.]$ is the probability of occurrence, and  $\left \lfloor{.}\right \rfloor$ represents the floor function. For readers' convenience, the most essential variables used throughout the paper are listed in Table  \ref {tab:vari}. {Furthermore, a list of key abbreviations and acronyms used in this paper can be found in Table  \ref{tab:abbr}.}

\begin{table}[]
\scriptsize
\renewcommand{\arraystretch}{1.3}
\caption{{Most  important variables used throughout the paper} }\vspace{-2mm}
\label{tab:vari}
\centering
\begin{tabular}{cclll}
\cline{1-2}
\multicolumn{1}{|c|} {Variable} & \multicolumn{1}{c|}{Definition }     &                 \\ \cline{1-2} \cline{1-2}
\multicolumn{1}{|c|}{$t_{su}$} & \multicolumn{1}{c|}{start-up time of cellular module }                                                               &  &  &  \\ \cline{1-2}
\multicolumn{1}{|c|}{$t_{pd}$} & \multicolumn{1}{c|}{power-down time of cellular module }                                                               &  &  & \\ \cline{1-2}
\multicolumn{1}{|c|}{  $t_{sync}$ } & \multicolumn{1}{c|}{synchronization time of cellular module}                                                               &  &  &  \\ \cline{1-2}
\multicolumn{1}{|c|}{ $\text{PW}_{\text{sleep}}$ } & \multicolumn{1}{c|}{ power consumption of cellular subsystem at sleep state }  
             &  &  &  \\ \cline{1-2}
\multicolumn{1}{|c|}{ $\text{PW}_{\text{active}}$ } & \multicolumn{1}{c|}{ power consumption of cellular subsystem at active state }   &  &  &  \\ \cline{1-2}
\multicolumn{1}{|c|}{ $t_c$ } & \multicolumn{1}{c|}{wake-up cycle }   &  &  & 
 \\ \cline{1-2}
 \multicolumn{1}{|c|}{ $t_{on}$ } & \multicolumn{1}{c|}{monitoring duration of WRx per w-cycle }   &  &  & 
 \\ \cline{1-2}
  \multicolumn{1}{|c|}{ $t_{sl}$ } & \multicolumn{1}{c|}{sleep period of WRx per w-cycle }   &  &  & 
 \\ \cline{1-2}
  \multicolumn{1}{|c|}{ $t_{of}$ } & \multicolumn{1}{c|}{time offset of wake-up scheme }   &  &  & 
 \\ \cline{1-2}
   \multicolumn{1}{|c|}{ $T_{ON}$ } & \multicolumn{1}{c|}{on-time duration}   &  &  & 
 \\ \cline{1-2}
   \multicolumn{1}{|c|}{ $T_{I}$ } & \multicolumn{1}{c|}{inactivity timer }   &  &  & 
 \\ \cline{1-2}
      \multicolumn{1}{|c|}{ $N_{w}$ } & \multicolumn{1}{c|}{wake-up timer}   &  &  & 
 \\ \cline{1-2}
    \multicolumn{1}{|c|}{ $P_{fa}$ } & \multicolumn{1}{c|}{probability of false alarm}   &  &  & 
 \\ \cline{1-2}
    \multicolumn{1}{|c|}{ $P_{md}$ } & \multicolumn{1}{c|}{probability of misdetection}   &  &  & 
 \\ \cline{1-2}
 \multicolumn{1}{|c|}{ $N_{cp}$} & \multicolumn{1}{c|}{ the cyclic-prefix  length   of the OFDM   symbol in terms of samples}                                                               &  &  &  \\ \cline{1-2}
\multicolumn{1}{|c|}{ $N_{b}$} & \multicolumn{1}{c|}{ the number of
samples of body of OFDM symbol}         
      &  &  &  \\ \cline{1-2}
\multicolumn{1}{|c|}{ $N_{g}$} & \multicolumn{1}{c|}{ the overall number of
guard subcarriers around of each PDWCH group}  &  &  &  \\ \cline{1-2}
\multicolumn{1}{|c|}{ $K$ } & \multicolumn{1}{c|}{length of  Zadoff$\mbox{-}$Chu     sequence }                                                               &  &  &  \\ \cline{1-2}
\multicolumn{1}{|c|}{ $M$} & \multicolumn{1}{c|}{ number of mobile devices within PDWCH group}                                                               &  &  &  \\ \cline{1-2}
\multicolumn{1}{|c|}{   $N$ } & \multicolumn{1}{c|}{size of FFT }                                                               &  &  &  \\ \cline{1-2}
\multicolumn{1}{|c|}{$i[m]$} & \multicolumn{1}{c|}{ wake-up indicator of $m^{th}$ mobile device }                                                               &  &  &  \\ \cline{1-2}
\multicolumn{1}{|c|}{$\tau[m]$} & \multicolumn{1}{c|}{cyclic-shift of $m^{th}$ mobile device  }                                                               &  &  &  \\ \cline{1-2}
\multicolumn{1}{|c|}{$Z_m[k]$} & \multicolumn{1}{c|}{DFT of root ZC sequence with cyclic-shift of $\tau[m]$ }                                                               &  &  &  \\ \cline{1-2}
\multicolumn{1}{|c|}{$R_q[k]$} & \multicolumn{1}{c|}{$N$-point FFT  output of  $q^{th}$ OFDM symbol}                                                             &  &  &  \\ \cline{1-2}
\multicolumn{1}{|c|}{ $\epsilon_f$ } & \multicolumn{1}{c|}{  fractional carrier frequency offset }                                                               &  &  &  \\ \cline{1-2}
\multicolumn{1}{|c|}{ $\epsilon_i$ } & \multicolumn{1}{c|}{  integer carrier frequency offset }                                                               &  &  &  \\ \cline{1-2}
\multicolumn{1}{|c|}{ $\delta$ } & \multicolumn{1}{c|}{integer-valued OFDM symbol's STO in terms of sampling instant time }                                                               &  &  &  \\ \cline{1-2}
\multicolumn{1}{|c|}{  $s$ } & \multicolumn{1}{c|}{ index of OFDM symbol carrying PDWCH  }                                                               &  &  &  \\ \cline{1-2}
\multicolumn{1}{|c|}{   $x$ } & \multicolumn{1}{c|}{\begin{tabular}[c]{@{}c@{}}  maximum number of consecutive OFDM symbols \\ within which PDWCH can be located \end{tabular}} &  &  &  \\ \cline{1-2}
\multicolumn{1}{|c|}{   $a$ } & \multicolumn{1}{c|}{\begin{tabular}[c]{@{}c@{}}  maximum number of subcarriers in integer CFO estimation \end{tabular}} &  &  &  \\ \cline{1-2}
\multicolumn{1}{|c|}{  $\Psi_q$ } & \multicolumn{1}{c|}{ PDP of the received signal at $q^{th}$ OFDM    symbol }                                                               &  &  &  \\ \cline{1-2}
\multicolumn{1}{|c|}{   $\psi_q$} & \multicolumn{1}{c|}{\begin{tabular}[c]{@{}c@{}} discrete periodic correlation function     of    received signal and complex    \\ conjugate of frequency-shifted version of the root zadoff$\mbox{-}$chu sequence\end{tabular}} &  &  &  \\ \cline{1-2}
\multicolumn{1}{|c|}{   $E_q$ } & \multicolumn{1}{c|}{\begin{tabular}[c]{@{}c@{}}  received  energy within the sliding window at $q^{th}$  OFDM symbol \end{tabular}} &  &  &  \\ \cline{1-2}
\multicolumn{1}{|c|}{ $t_{s}$ } & \multicolumn{1}{c|}{ session inter-arrival time }                                                               &  &  &  \\ \cline{1-2}
\multicolumn{1}{|c|}{ $t_{pc}$ } & \multicolumn{1}{c|}{ packet call  inter-arrival time within the session }                                                               &  &  &  \\ \cline{1-2}
\multicolumn{1}{|c|}{ $t_{p}$ } & \multicolumn{1}{c|}{ packet inter-arrival time within the packet call }    
&  &  &  \\ \cline{1-2}

\multicolumn{1}{|c|}{$\text{S}_k$ } & \multicolumn{1}{c|}{ the $k^{th}$ power state of NM }   
&  &  &  \\ \cline{1-2}
\multicolumn{1}{|c|}{ $\text{P}_k$ } & \multicolumn{1}{c|}{ steady state probability that
NM is at $\text{S}_k$}   
&  &  &  \\ \cline{1-2}
\multicolumn{1}{|c|}{ $\text{P}_{kl}$ } & \multicolumn{1}{c|}{ transition probability from $\text{S}_k$ to $\text{S}_l$ } 
&  &  &  \\ \cline{1-2}
\multicolumn{1}{|c|}{ $\omega_k$ } & \multicolumn{1}{c|}{holding times for $\text{S}_k$}   
&  &  &  \\ \cline{1-2}\multicolumn{1}{l}{}    & \multicolumn{1}{l}{}                                                                  &  &  & 
\end{tabular}
\end{table}

 \begin{table}[]
\scriptsize
\renewcommand{\arraystretch}{1.3}
\caption{{Key abbreviations and acronyms used throughout the paper} }\vspace{-2mm}
\label{tab:abbr}
\centering
\begin{tabular}{cclll}
\cline{1-2}
\multicolumn{1}{|c|} {Abbreviation/Acronym} & \multicolumn{1}{c|}{Definition}     &                 \\ \cline{1-2}
\cline{1-2}
\multicolumn{1}{|c|} {ASIC} & \multicolumn{1}{c|}{application-specific integrated circuit}     &                 \\ \cline{1-2}
\multicolumn{1}{|c|} {BBU} & \multicolumn{1}{c|}{baseband processing unit}     &                 \\ \cline{1-2}
\multicolumn{1}{|c|} {CFO/STO} & \multicolumn{1}{c|}{carrier
frequency/symbol time offset}     &                 \\ \cline{1-2}
\multicolumn{1}{|c|} {DRX} & \multicolumn{1}{c|}{discontinuous  reception}     &                 \\ \cline{1-2}

\multicolumn{1}{|c|} {gNB} & \multicolumn{1}{c|}{5G base-station}     &                 \\ \cline{1-2}
\multicolumn{1}{|c|}{LTE} & \multicolumn{1}{c|}{long-term evolution  }                                                               &  &  &  \\ \cline{1-2}
\multicolumn{1}{|c|} {LPO/HPO} & \multicolumn{1}{c|}{  low/high-power  low/high-precision
electronic oscillator
}     &                 \\ \cline{1-2}
\multicolumn{1}{|c|} {NM} & \multicolumn{1}{c|}{new  modem}     &                 \\ \cline{1-2}
\multicolumn{1}{|c|} {PDP} & \multicolumn{1}{c|}{power  delay
profile  
}     &                 \\ \cline{1-2}
\multicolumn{1}{|c|} {PDCCH/PDSCH} & \multicolumn{1}{c|}{physical downlink control/shared channel}     &                 \\ \cline{1-2}
\multicolumn{1}{|c|} {PDWCH} & \multicolumn{1}{c|}{ physical downlink wake-up channel}     &                 \\ \cline{1-2}
\multicolumn{1}{|c|} {RFIC} & \multicolumn{1}{c|}{radio frequency integrated circuit}     &                 \\ \cline{1-2}
\multicolumn{1}{|c|} {UE} & \multicolumn{1}{c|}{user equipment}     &                 \\ \cline{1-2}
\multicolumn{1}{|c|} {WRx} & \multicolumn{1}{c|}{wake-up  receiver }     &                 \\ \cline{1-2}
\multicolumn{1}{|c|} {WSN} & \multicolumn{1}{c|}{wireless  sensor  network}     &                 \\ \cline{1-2}
\multicolumn{1}{|c|} {WI} & \multicolumn{1}{c|}{ wake-up indicator }     &                 \\ \cline{1-2}
\multicolumn{1}{|c|} {w-cycle} & \multicolumn{1}{c|}{wake-up cycle}     &                 \\ \cline{1-2}
\multicolumn{1}{|c|} {ZC} & \multicolumn{1}{c|}{zadoff$\mbox{-}$chu}     &                 \\ \cline{1-2}
\multicolumn{1}{l}{}    & \multicolumn{1}{l}{}                                                                  &  &  & 
\end{tabular}
\end{table}

 The rest of this article is organized as follows. Section~\ref{section:Motivation}  provides the problem description and motivation to solve it. Section~\ref{section:sota} summarizes the current state of wake-up schemes  available in  literature.  Section~\ref{section:wus} introduces and describes the proposed novel wake-up scheme.The proposed solutions for WRx detection and synchronization processing are formulated in Section~\ref{section:system}.  Section \ref{Markov} analyzes the average latency and power consumption of the proposed wake-up scheme. Section~\ref{h1} briefly addresses the  implementation aspects of the WRx, and provides measurement-based power consumption model for the NM.     Section~\ref{section:simulation} presents the numerical results to validate the proposed scheme. Finally, Section~\ref{section:conclusions} concludes the article, {while further implementation specifics are provided in the Appendix as supplemental material.}

\section{Motivation and Problem Description}
\label{section:Motivation}
In general, the $5$G networks are expected  to  provide high data rates  in order to deliver diverse set of new  services such as ultra-high-definition video streaming or augmented reality to growing number of data-hungry users \cite{Boccardi2014,Alamouti}. To satisfy the aggressive requirements of such services, advanced signal processing techniques and high radio access bandwidths, extending up to $100$ MHz and $400$ MHz at sub-$6$ GHz and mmWave spectrum bands are imperative \cite{nr2017,soheil2017}. Over the last decade, the trend for increased bandwidths has been following the same exponential increase as Moore's Law for semiconductors \cite{Alamouti}. However, higher bandwidth communication consumes considerably higher power, and can shorten the mobile devices battery lifetime quickly. Due to the fact that  the battery evolution is not as mature as   the advances in semiconductor industry  \cite{Lauridsenthesis},  power saving methods are vital to extend $5$G mobile devices functionalities and usability.

Traditional application  layer protocols such as HTTP and FTP     generate  traffic only during the connected mode, and once the radio resource control (RRC) inactivity timer has expired, the device moves to an idle mode \cite{Maruti2013}. However, emerging mobile Internet applications using  XMPP  (such as Facebook and Twitter) generate a constant stream of autonomous or user generated traffic at all times,    leading to frequent back and forth transitions between connected  and idle modes, while sending mostly short bursts of data. This easily drains the device battery and causes excessive signaling overhead in the networks  \cite{Maruti2013}. Therefore, the DRX mechanism has been introduced already in LTE, in connected mode, in  order  to monitor the  physical downlink control channel (PDCCH) only during the {{active}} periods within either short DRX cycle (with maximum     of $512$ ms) or long DRX cycle  (with maximum    of $2560$ ms), and to   switch  off   some components of cellular subsystem in the {{sleep}} period. Eventually, the mobile device can reside most of the time in connected mode, while use DRX to save energy \cite{Maruti2013}.

\begin{figure}[t!]
\centering
\includegraphics[scale=.9]{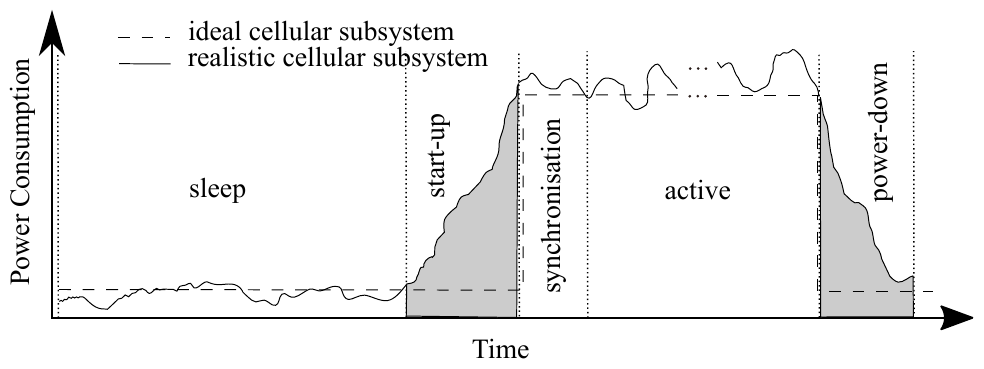}
\caption{Ideal and realistic power consumption  of cellular subsystem in long DRX. In case of short DRX, start-up and power-down time are shorter, and the required power for sleep period is higher. In addition, synchronization stage is not required, since the device is in-synchronization during sleep period of short DRX.}
\label{fig:realisticdrx}
\end{figure}

 Ideally, the cellular subsystem is expected to transit from  {{sleep}}  to  {{active}}  state or vice versa sharply, as illustrated in dashed line in Fig.~\ref{fig:realisticdrx}. However, because of  hardware preparation time, cellular subsystem power consumption shape     changes from purely ideal rectangular to a smooth contour without sharp edges as shown in solid line in Fig.~\ref{fig:realisticdrx}. As a result, considerable amount of energy is wasted during start-up and power-down periods of a DRX cycle, as depicted in gray area in Fig.~\ref{fig:realisticdrx}. The smoother power consumption profile results into a  lower {sleep} time, than the ideal scenario, consequently reducing sleep ratio  and increasing energy consumption of DRX.

 Typical  cellular subsystem is comprised of sequential electronic blocks, where each building block's functionality and availability are dependent on the previous blocks,  posing an additional start-up time ($t_{su}$) and power-down time ($t_{pd}$).  Therefore, in   real settings during DRX,   the  cellular subsystem operates in five different phases,   described as follows while being also marked in Fig.~\ref{fig:realisticdrx}:

\begin{itemize}
\item 
 {\textbf{{Sleep}}}: in this phase, some of the hardware components in the cellular subsystem   are either completely switched-off or operated in
low-power modes.   The exact components to be switched off are vendor-specific, and depend on DRX-cycle length. Generally, once in long DRX sleep (also known as deep sleep),  the cellular subsystem keeps track of the number of frames by means of a low-power low-precision electronic oscillator (LPO); 

\item 
\textbf{{Start-up}}:
during transition from deep sleep to active state, the high-power high-precision oscillator (HPO) is switched-on. However, the HPO requires some time to settle and  reach its optimum functionality, causing prolonged start-up time, and hence leading  to the  gradual rise in  power consumption profile. Furthermore, it is paramount that the baseband (BB)  synthesizer and the RF phase locked loop (PLL)  are  phase-matched   \cite{Lauridsen2014}. Therefore, in addition to the switch-on delay of the HPO, also the PLL lock time contributes to the start-up time;

  \item 
\textbf{{Synchronization}}: to assure that mobile device is in synchronization with the network, the mobile device needs to demodulate the synchronization signals, which are sent periodically (in case of LTE, every  $5$ ms; configurable in {5G new radio (NR)}). Meantime, the device can decode the system information, and also estimate the channel. The synchronization delay depends vastly on frame structure;	

\item 
{\textbf{{Active}}}:  in this phase, all hardware components  in the cellular subsystem are fully switched-{on}, and synchronized, ready to  receive PDCCH and physical downlink shared channel (PDSCH);

\item 
\textbf{{Power-down}}: powering down of integrated circuits of the BBU does not translate to the sharp turn-off, in other words, the cellular subsystem still consumes some power when turning off. 
\end{itemize}

As a concrete example, Table \ref{drxpower} expresses the average transitional times and power consumption values of both short and long DRX cycles based on  measurements of multiple different LTE cellular subsystems \cite{Lauridsenthesis}. As can be seen, the sleep power consumption ($\text{PW}_{\text{sleep}}$) is much higher in the short DRX cycle than in long one.   This, in turn, has a  direct influence on $t_{su}$ and $t_{pd}$. In case of short DRX, due to fact that BBU is  ON  during the sleep period, $t_{su}$ and $t_{pd}$ are  shorter than in long DRX. In short DRX cycle, the time required for synchronization, $t_{sync}$, is negligible. Required power for active state ($\text{PW}_{\text{active}}$)  varies based on IC/modem design, bandwidth, and other communication parameters.

\begin{table}[t!]
\small
\renewcommand{\arraystretch}{1.3}
\caption{Average transitional times  and power consumption values  of LTE cellular subsystem during short and long DRX when carrier bandwidth is $20$ MHz \cite{Lauridsenthesis}}
\label{drxpower}   
\centering
\scriptsize
\begin{tabular}{|c|c|c|c|c|c|c|c|}
    \hline
     DRX Cycle &$\text{PW}_{\text{sleep}}$& $t_{su}$& $\text{PW}_{\text{active}}$ &  $t_{sync}$&   $t_{pd}$  \\
    \hline
    \hline
      short&  $395$ mW&   $\leq 1$ ms&   $850$ mW&   $0$ ms &   $\leq 1$ ms\\
          \hline
      long&  $9.8$ mW&   $15$ ms&   $850$ mW&   $10$ ms &   $10$ ms\\
          \hline
\end{tabular}
\end{table}

With respect to  latency requirements, it is beneficial to process     PDCCH  in a very short DRX cycle to receive  uplink (UL)  grants or downlink (DL) data traffic, and promptly make an appropriate reaction. However,  decoding the \mbox{PDCCH}   requires a fast Fourier transform (FFT) whose size depends on the carrier bandwidth, and employs a blind decoding approach, where it hypothesizes over $44$ options of PDCCH locations \cite{trxxxx,trxxxx2}. Especially for higher bandwidth carriers,   the PDCCH rendering is very computationally intensive and power consuming. Therefore, {decoding} PDCCH  frequently reduces the advantages of DRX,  and high power consumption overhead is inevitable.

 It has also been shown in   \cite{youtube}  that   the time period, that mobile device monitors the channel  without any data allocation has a major impact on  battery energy consumption. For instance, according to the experimental results, video streaming and web browsing traffic as the  representative use cases have unscheduled PDCCH  for $25$\%  and  $40$\% of time, respectively \cite{youtube}. Furthermore, this problem is severe in unsaturated traffic scenarios, where many of the DRX cycles have eventually no data allocation for a particular mobile device. In general, there is also a trade-off between energy consumption and latency. That is, to reduce energy consumption, the DRX cycle needs to be made longer, which in turn leads to higher latency. 
 
 In this article, motivated by above, we propose a novel  wake-up scheme as an efficient  method to tackle the trade-off between latency and battery lifetime by reducing the power overhead of the DRX procedure, caused by unscheduled DRX cycles, while still maintaining low latency.

\section{Wake-up Methods State of the Art}
 \label{section:sota} 
The related wake-up scheme works    can be classified based on the    wireless communication system (WSN, WLAN and cellular), for which the wake-up scheme is been developed. The original works on the wake-up methods are mainly focusing on WSNs and IoT-oriented WLANs, where the energy constraints of the end nodes are very tight. In spite of    the promising foreseen evolution   of the wake-up scheme in  cellular communication systems,  the wake-up concept needs to be studied further for the latter, and has started to raise interest recently also in 3GPP standardization. In this section,   based on the existing works and documentations,  the state-of-the-art is reviewed briefly.
  
 \subsection{WSN and WLAN Systems} 
 Authors in \cite{lont2009,Zhang2009,lin2004, Demirkol2009}, developed different wake-up schemes. Previous works on WRx-empowered WSNs can be classified into the continuous  and duty cycled channel monitoring. Continuous monitoring scenario prominently tackles use cases, where ultra-low latency of wireless nodes is the major requirement \cite{lont2009,Zhang2009}. However, in case of infrequent packets, energy consumption of the  wireless node is remarkably high, and WRx can drain the battery energy if operating on continuous basis.  Recently,    duty-cycled ultra low-power WRx  for WSNs has been developed \cite{Mazloum2014}, building on  periodic wake-up beacons, which are transmitted ahead of data packets for synchronization of the communicating nodes.  Mazloum \emph{et al.} \cite{Mazloum2014} also calculated  closed-form expressions for the corresponding detection and false-alarm probabilities.

  Demirkol \textit{et al} \cite{Demirkol2009} provided a comprehensive overview and insight into WRx, and investigated the benefits achieved with   WRx along with the challenges observed in WSNs. In addition, they presented an overview of state-of-the-art hardware and networking protocol proposals as well as classification of WRx schemes. Moreover,  authors in \cite{wpwrx} introduced the concept of wireless-powered wake-up receiver,  reducing the energy consumption of the wireless node considerably. The proposed  receiver scavenges the RF energy from the received signal to power its sensor, communication and processing blocks. The proposed scheme can be utilized for a wide range of energy-constrained wireless applications such as wireless sensor actuator networks and  machine-to-machine communications.

On WLAN  system standardization side, IEEE 802.11 working group    is currently considering WRx as a companion radio for the main receiver \cite{eee1}.  The major chip and wireless players such as   Intel, MediaTek, and Ericsson  have presented technical contributions to include WRx in IoT use cases, consistently  \cite{eee2,eee3,eee4}. It is expected that WRx operation will be fully standardized for IEEE 802.11 during 2020. In advocated proposal, WRx receives and decodes a wake-up packet without any help from the main receiver \cite{eee5}. It has a cascaded wake-up signalling structure to support multiple stations. Furthermore, payload of wake-up packet is modulated with on-off keying because of its simplicity, although it is known to be prone to noise \cite{eee5}.

\subsection{Cellular Systems} 
While WRx concepts have been extensively studied in WSN context, there are only a very few  works  available for scheduled frame-based cellular communication systems. Very recently, \cite{Lauridsenthesis, Lauridsen2016} showed that the concept of WRx, when adapted to   cellular communication, can provide $90$\%  lower energy consumption. Authors in  \cite{Lauridsenthesis} combined microsleep, DRX and discontinuous transmission, and an aperiodic sleep mode to enhance  battery lifetime of $5$G mobile devices. However, the  wake-up signaling structure and associated processing solutions are not addressed.  Recently, the authors in \cite{pregrant} and \cite{8616818}, enhanced the  DRX   by enabling 5G mobile device to read small  pre-grant message during active time, instead of executing full PDCCH processing. The preliminary numerical results show that such a scheme can reduce power consumption of mobile devices considerably, at the cost of negligible increments in signaling overhead.

{Authors in \cite{Ponna} developed a wake-up scheme for machine-type communication (MTC) with idle mode emphasis. The approach builds on a separate standalone receiver for wake-up signal processing, while mathematical analysis and system parameter optimization were not pursued. The reported achievable misdetection rate is ca. 2\% while the corresponding false alarm rate is 5.3\%. The study and the proposed scheme did not address synchronization aspects nor synchronization assistance to the BBU. }

{\color{black} Wake-up based solutions have started to raise interest recently also in 3GPP mobile network standardization. Specifically, a wake-up scheme to optimize the LTE-based enhanced-MTC (eMTC) and narrow-band IoT (NB-IoT) deployments were recently introduced in Rel. 15 \cite{TS36.300}. The assumed wake-up concept is based on a narrowband indicator signal,  transmitted over  the available symbols of the configured subframes. It conveys $504$ unique cell identities, as per the narrowband secondary synchronization signal (SSS). Furthermore, the wake-up signal is  scrambled by cell-specific code for mitigation of inter-cell interference \cite{TS36.211}. The proposed wake up scheme for eMTC and NB-IoT  is applied in idle mode before monitoring paging signaling and optimized for small infrequent packet transmissions ranging  between some $50$ to $200$ bytes, a few times per day, and the wake-up scheme parameters are configured by higher layers \cite{trxxxx}. Due to the promising power saving efficiency of such wake-up concept, it is also considered as the starting point in some 5G NR power saving studies, as discussed in \cite{NR_PS}. Generally, because 5G NR supports very high speed data rates, the user plane data profile tends to be bursty and served in very short duration, and hence wake-up scheme can trigger user equipment (UE) for network access from power saving mode.   Hence, the work and methods in this manuscript are very well inline with the upcoming 5G NR standardization efforts, and can be seen as an attempt to define, explore and analyze a concrete wake-up scheme towards the future releases of 5G NR, basically applicable for idle, inactive and connected modes.}

  {Finally, it is noted that the original wake-up concepts in the WSN context are commonly based  on unlicensed and contention-based channel access, while cellular communication systems build on scheduled frame-based access. Additionally, the processing capabilities, cost and power consumption of WSN nodes, overall, are commonly much lower than those of the cellular devices, especially mobile broadband capable UEs. Thus, also the available WRx power budget and assumed processing capabilities are substantially lower in WSN context, in general. As optimizing the power consumption is the highest priority in WSN context, one may also argue that the corresponding WRx solutions may even deactivate some of the receiver active components, such as the low noise amplifier (LNA). This leads to power consumption values in the range of a few $\mu$W, compared to the WRx power consumption range of a few tens of mW in the cellular context \cite{Lauridsenthesis}. LNA deactivation together with other hardware issues easily lead to substantially reduced WRx sensitivity, compared to the sensitivity requirements of cellular devices \cite{Demirkol2009}.  Finally, the observation time of the wake-up signal in WSN-related wake-up designs, commonly ranging in few tens of milliseconds, is substantially longer than the available observation time for cellular communication which is in the range of sub-milliseconds. The long observation time increases, among others, the buffering delay of the wake-up scheme.}

\section{Proposed Wake-up Scheme}
\label{section:wus}
\subsection{Operating Principle at High Level}
 In general, depending whether an RRC connection is established or not, the proposed wake-up scheme described in this article can be adopted for idle, inactive or connected modes. However, for presentation  brevity, we mainly focus on the connected mode in the following.Independent of the proposed wake-up scheme, successful completion of the cell search and selection procedures as well as acquiring initial system information and synchronization are required.  The synchronization procedure occurs, in general, in two stages. In the first stage, the device acquires symbol timing, frequency offset, and cell ID using the primary synchronization signal (PSS). Then, in the second stage, the mobile device detects the frame boundary, physical cell group, and the cyclic prefix (CP) length by using the SSS \cite{nr2017}.

Each mobile device in the connected mode  operates in either the power-active    or the power-saving mode.  In the former,  the device can receive or transmit in a continuous manner, which is appropriate, e.g.,  for ultra-reliable low latency communication (URLLC) purposes. In the latter, which is the focus of this work, the mobile device is configured with the proposed wake-up signaling and associated WRx in order to save battery lifetime, suitable for eMBB services as well as for massive machine-type communication (mMTC) services, which commonly contain small amounts of data at short intervals.

In the power-saving mode, the overall proposed wake-up scheme works as follows. Within a wake-up cycle, referred to as the \emph{w-cycle} in the continuation with length denoted by $t_{c}$, the WRx monitors the so-called  physical downlink wake-up channel (PDWCH) for a specific on-duration time ($t_{on}$) to determine if any data is scheduled for it or not. Occasionally, based on either expiration of the configurable wake-up timer or interrupt signal from WRx, the BBU switches on, and thereon may (re)acquire synchronization, decodes both  PDCCH  and  PDSCH, and  performs connected-mode procedures such as reporting channel state and neighboring cell measurements to  the network.

Because of the very simple hardware architecture of WRx, as discussed more extensively in Section~\ref{h1}, the start-up and power-down periods for WRx are extremely short. As elaborated more in the next subsection and Section \ref{section:system}, the processing requires demodulation of the specific subcarriers of only one or few OFDM symbols in each w-cycle, with overall duration of $t_{on}$, in contrast to normal DRX, where BBU   needs to operate full bandwidth for multiple symbols.       In other words, narrowband reception of wake-up signaling  requires much less signal processing, consequently needing also less memory and operations. Moreover, its narrowband signal structure, addressed in more details in the following subsection, has improved sensitivity due to its low in-band receive noise. Thus, it can be effectively applied and processed even in the presence of severe path loss and multipath propagation.

The length of the w-cycle, $t_{c}$, can also be flexibly reconfigured and thus shortened, at the cost of reasonable extra energy consumption, and thus the adoption of the proposed wake-up scheme and the NM can also facilitate low communication latencies when properly configured.  Additionally,  built-in self-synchronizing signal structure and the utilization of the HPO remove the need for separate  synchronization stage for WRx. {Moreover, WRx can assist the BBU to obtain synchronization immediately, leading to further reduction in energy consumption and latency  (i.e., $t_{sync}\approx 0$, see Fig. \ref{fig:realisticdrx} and Table \ref{drxpower}).}

\begin{figure}[t!]
\centering
\includegraphics[scale=.92]{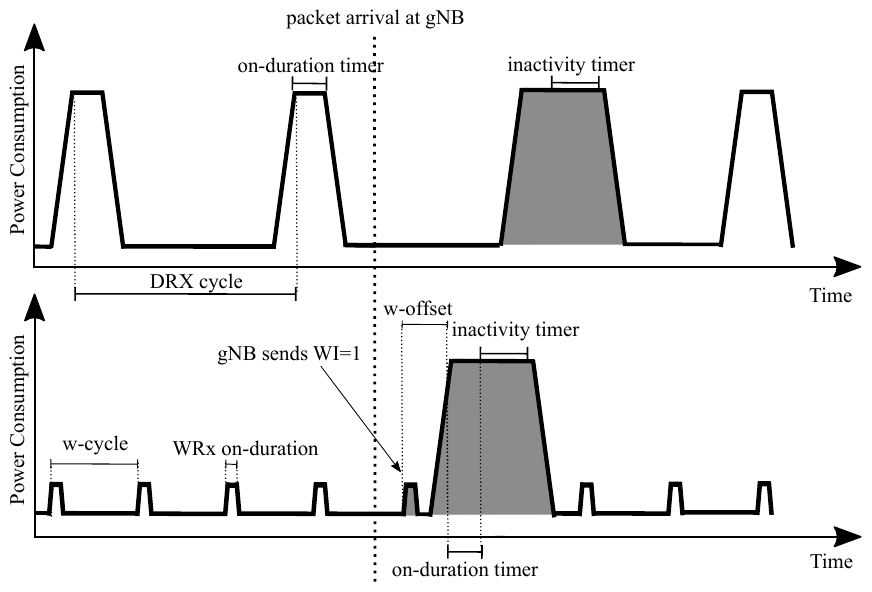}
\caption{Principal operation and corresponding power consumption characteristics of the ordinary DRX mechanism (upper figure) and the NM utilizing the  proposed wake-up scheme (lower figure). }
\label{fig:drxvswrx}
\end{figure}

In order  to identify when to wake up the BBU, the WRx  monitors the  PDWCH  every w-cycle.
The wake-up signaling structure transmitted every w-cycle is a unique  code sequence, sent as a wake-up indicator (WI) for a mobile device that needs to wake up. {Once there is packet arrival from core network to gNB, the gNB sends WI=1 to   UE in upcoming wake-up instance.}   When such a WI  has been sent to WRx, the network expects the corresponding mobile device to decode the PDCCH with time offset of $t_{of}$. In other words, PDWCH indirectly informs  WRx   of potential scheduling on PDCCH with time offset of $t_{of}$ ($t_{of}\ge t_{su}$) or rather if it can skip interrupting BBU for the rest of $t_{c}$.  {In order to improve the reliability of the wake-up scheme, once WI has been sent, the network will send additional   WIs to the mobile device every   w-cycle  until   the network receives proper ACK message. Similarly, the WRx can basically carry on the reception of PDWCH, according to the prevailing w-cycle pattern, after identifying WI=1 until receiving the target PDCCH message.}  Fig. \ref{fig:drxvswrx} depicts the operation and power consumption of the proposed NM at principal level, while also illustrating the conventional DRX-enabled cellular subsystem operation for reference.

After receiving the PDCCH message at active state for on-time duration of $T_{ON}$, BBU initiates its inactivity timer for duration  of $T_{I}$. After the inactivity timer is initiated, and if a new PDCCH message is received before the expiration of $T_{I}$, BBU re-initiates its inactivity timer. However, if there is no PDCCH message received  before expiration of the inactivity timer, a sleep period starts, and NM switches to its sleep state, while WRx operates according to its w-cycle.

Due to the presence of noise, interference, distortion or  mis-synchronization over PDWCH,  two different WI errors, namely  misdetection (wrongly detecting WI as $0$) and false alarm (wrongly detecting WI as $1$),  are inevitable. False alarm can cause unnecessary power consumption of the BBU, while misdetection can add the buffering delay and wastes radio resources.  Therefore,   the requirements for both the probability of  misdetection   ($P_{md}$) and  the probability of false alarm ($P_{fa}$) are generally strict, in the order of $1-10$\%.  Simulation results in Section \ref{section:simulation},   verify that the proposed scheme, building on the signaling structure and decoder principle described in Section \ref{section:system}, can achieve very low $P_{md}$ and $P_{fa}$  for SNRs even below $0$ dB.

{It is additionally noted that in the proposed wake-up scheme, if WRx does not detect WI for $N_w$ consequent w-cycles, referred to as the wake-up timer, it is assumed that the BBU will be switched on to perform connected mode procedures in order to eliminate the scenario that WRx is out of synchronization. While doing so, the UE can also report channel state information, as well as to perform general mobility procedures based on needs. The value of the wake-up timer can be configured by the network at device level based on, e.g., UE category, UE maximum mobility, traffic type and/or type of the network deployment. For instance, in case of ultra-dense networks,   handovers may happen very frequently, thus the value of the wake-up timer can be configured to be fairly small to make sure mobility management is not compromised. Additionally, 5G systems should support also very high mobility levels, thus as shown in \cite{mobility}, UE channel measurements need to be reported fairly often, otherwise handover failure rate increases largely. Properly set wake-up timer removes thus the possibility that UE is forced to interrupt its ongoing connection due to exiting the serving cell coverage area and entering the new cell without handover procedures, which would lead to poor QoS and executing computationally-expensive and time-consuming tasks of cell selection and camping. At an other extreme, with e.g. zero-mobility IoT sensors, the value of $N_w$ can be set substantially larger to optimize the energy saving, within the limits of feasible synchronization.   In our work, we prioritize the support for high-mobility UEs and assume a practical example wake-up timer value of $600$ ms (i.e. $N_w\leq \left \lfloor{600/t_c}\right \rfloor$). We emphasize that this is just one example numerical selection, and in practice the value of the wake-up timer can be configured by the network.}

\subsection{Frame and Signal Structures}
\label{frame}
In this work, overall and in the numerical evaluations,  we consider LTE and 5G NR like frame structure, where transmission over the subcarriers is arranged into radio frames of $10$ ms long, each of which is divided into ten equally-sized individual subframes. Furthermore, we assume that each subframe is consisting of $14$ consecutive OFDM symbols   while the subcarrier spacing is assumed to be $15$ kHz \cite{nr2017}. The OFDM symbol has $N_{cp}+N_{b}$ samples, where   $N_{cp}$ is the CP length   of the   symbol,  and $N_{b}$ is the number of samples of the body of the OFDM symbol. Additionally, within each frame,  PSS and SSS are used for synchronization purposes. In this work,  the introduced wake-up scheme is adopted for the aforementioned  frame structure, in the context of frequency-division duplexing (FDD) networks, while it can also be generalized to other potential frame structures easily.

The proposed signaling is based on the $5$G base station, commonly called gNB,  transmitting a set of Zadoff$\mbox{-}$Chu (ZC) signatures with different cyclic shifts over a dedicated and pre-reported set of subcarriers, carrying  set of  WIs     along with   synchronization. A pool of known  cyclic shift sequences is allocated to each gNB within a cell with a cell-specific root index, providing low inter-cell interference.  The  ZC sequences are known to have ideal cyclic auto-correlation, which is important for obtaining an accurate timing estimation and WRx identification. Additionally, the cross-correlation between different sequences based on cyclic shifts of the same  ZC root sequence is zero  as long as the cyclic shift used when generating the sequences is larger than the maximum DL    propagation time in the cell plus the maximum delay spread of the channel \cite{erik}.

The root ZC sequence, denoted by $z^{r}[n]$, is a polyphase  exponential ZC sequence     with root index of $r$, and can be formulated as
\begin{equation}
z^{r}[n]=\text{exp} \left \{{-j{\frac{ \pi r n (n+1) }{K}}} \right \}~\text{for}~~n\in \{0,...,K-1\},  
\end{equation}
 \noindent where  $K$ refers to the length of sequences,   assumed odd, and needs to be integer larger than and relatively prime with respect to  $r$ \cite{Guey}. An odd-length ZC sequence is symmetric to its center element, and enables design of  hardware-efficient   approach for its generation.  Further, if $K$ is prime, the discrete Fourier transform (DFT) of $z^{r}[n]$ is another ZC sequence, denoted as $Z^{r}[k]$.  In \cite{Popovic} closed-form expressions for the DFT of cyclically shifted ZC sequences of arbitrary length   are obtained.  For the sake of  readability, root index notation  is omitted in the rest of the paper.

PDWCH is transmitted within the first symbol of subframe corresponding to the w-cycle. The main reason for transmitting PDWCH in the first symbol is to provide adequate time ($\leq t_{of}$) for WRx to possibly switch on  BBU as early as possible. Therefore,  BBU can prepare to  decode PDCCH, and eventually, demodulate and decode    PDSCH. This reduces the processing delay, and thus the overall DL transmission delay. Furthermore,   to support full power transmission in order to have robust communication,   WIs are spread    over  adjacent subcarriers to reduce the power differences while at the same time providing the energy necessary for accurate reception also at cell-edge WRxs.  To fulfill this, the multiple  WIs are code multiplexed to a set of pre-defined subcarriers, which enables   efficient usage of radio resource elements, while also  reducing   the power consumption of the WRx. We refer to a set of WIs transmitted on the same set of subcarriers as a PDWCH group.

The WRx can  derive the resource elements in the control region where   PDWCH is mapped by reading   PDWCH group index value implying root index value of ZC sequence,  and     frequency offset parameter  (difference relative to be DC bin or channel center-frequency). Each   carrier or, equivalently    each cell may have  none, one or multiple PDWCH groups, adjusting to match the  PDWCH demand, depending on the number of overall WRxs and carrier bandwidth. For instance,  {in case of ultra-dense networks, where    small number of   connected mobile devices    may monitor   PDWCH   through their WRxs}, the required number of PDWCH group is    one,    and a larger part of the DL control can accommodate   L1/L2 control signaling.

The PDWCH parameters, i.e., the PDWCH group index  and the unique cyclic shift within group of a given WRx, as well as the parameters for determining the     frame number and subframe numbers in which the PDWCH  is available can be signaled by higher layers. For this purpose, RRC message can be used by the network to configure, activate and deactivate wake-up scheme and the corresponding parameters on a per device basis. In order to reduce signaling overhead,   the configuration parameters of the WRxs   can vary  {semi-statically}, and independently for each WRx, in order to absorb the  dynamics of the network. In the rest of this article, for presentation and notational simplicity, we assume that there is only one PDWCH group.

\subsection{RRC Modes and Signaling}
\begin{figure}[t!]
\centering
\includegraphics[scale=.47]{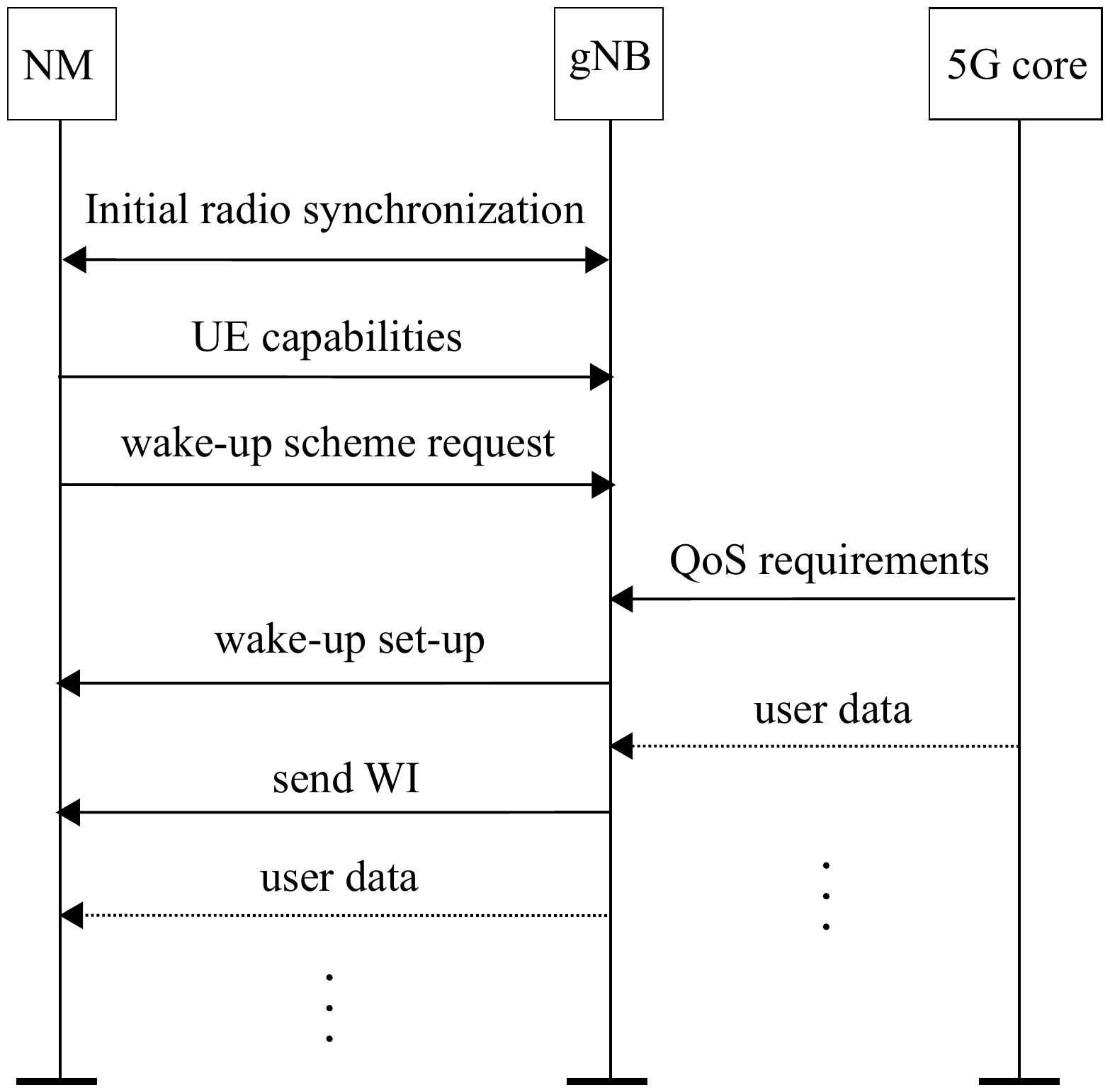}
\caption{{Illustration of the basic signaling involved for requesting wake-up as service and the corresponding wake-up scheme parameter setup. } }
\label{fig:5gcore}
\end{figure}

{In general, the wake-up based NM concept is agnostic to the RRC modes, and can thus be adopted for idle mode (where delay bound is in range of some hundreds of milliseconds), inactive mode (delay bound being in range of a few tens  of milliseconds \cite{Lauridsenthesis}) and the connected mode. The initial downlink timing synchronization is  required when the NM moves to power saving mode.  Once the NM has acquired the synchronization through PSS and SSS, the NM signals its capabilities ($t_{su}$ and $t_{pd}$) to the network. Then, the NM sends a wake-up scheme request as service to gNB and 5G core network. In response, QoS requirements (e.g., the maximum delay bound) and traffic pattern of ongoing connection are provided into gNB. Then, based on QoS requirements and other parameters (such as traffic arrival rate, delay bound, power profiles and start-up/power-down periods), the gNB configures the wake-up scheme parameters ($T_I$, $T_{ON}$, $t_c$, $K$,  $t_{of}$, the PDWCH group index  and the unique cyclic shift within group). After configuring the wake-up parameters, the gNB may send WI in pre-known wake-up instants, while the NM listens to the PDWCH with period of $t_c$. Once WI equals to one, the gNB sends user data and corresponding signaling (PDCCH and PDSCH) to the target UE with timing offset of  $t_{of}$.  Fig. \ref{fig:5gcore} illustrates the overall procedure and the involved essential signaling at connected mode.}

{In general, the DRX and the wake-up scheme   have similar procedures for parameter configuration. However, one of the main advantages of wake-up scheme over DRX is the need for reduced parameter re-configurations. For given delay requirements, NM can satisfy delay requirements with fixed parameter configuration for broad traffic variations, while DRX is very sensitive to traffic changes, and essentially requires re-configuration of its parameters frequently. The frequent re-configuration of   DRX parameters  to adapt to the traffic dynamics can, in turn, largely increase control signaling overhead, while also increases the UE energy consumption due to decoding the corresponding control signaling. By utilizing the wake-up scheme,  energy consumption of NM can be reduced with properly configured wake-up parameters without additional signaling overhead. Furthermore, for application areas such as MTC, where sensors can be aperiodically polled by either a user or a machine, the traffic will have very non-periodic patterns. For such scenarios, the DRX may not fit well, while the wake-up scheme has more suitable characteristics being more robust and agnostic to the traffic type. }

Finally, it is noted that since the PDWCH is mapped onto a set of $K$ adjacent subcarriers with the other subcarriers being used by DL control data symbols,   PDWCH should be extracted from   DL signal before the detection stage. This requires a filtering operation of the received signal,   increasing the hardware complexity to some extent, and can also lead to slight performance degradation since the contribution of the rest of the DL control subcarriers cannot be totally filtered out due to the spectral leakage. To alleviate this issue, $N_g$ additional subcarriers   in both    sides of PDWCH   are assumed to be muted.

\section{Wake-up Signal Detection}
 \label{section:system}
\subsection{Basics}
\begin{figure}[t!]
\centering
\includegraphics[scale=.75]{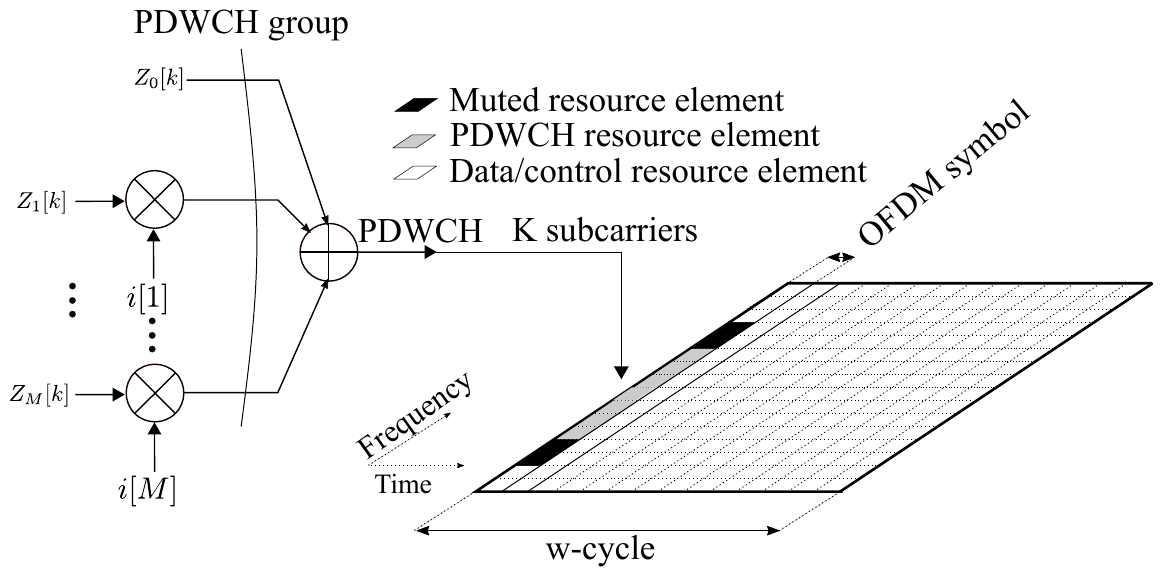}
\caption{Time-frequency grid illustrating PDWCH signal structure, $Z_m[k]$ refers to root ZC sequence with cyclic-shift of $\tau[m]$. }
\label{fig:structure}
\end{figure}

  We consider a multi-user scenario, where each sequence within PDWCH group is chosen from a set of $M$ cyclic-shifted ZC  sequences with length of $K$. Furthermore, during  first   OFDM symbol of predefined subframe,  $K$ contiguous subcarriers are utilized while the exact location of PDWCH is indicated by  the frequency offset parameter. For simplicity, we assume that subcarriers  with relative indices to the DC subcarrier  $k\in \{0,...,K-1\}$  are used for PDWCH. Therefore, the DFT of the PDWCH  group signal, denoted by $Y[k]$, covering $M$ mobile devices in the cell can be expressed as 
\begin{equation}
Y[k]=Z_0[k]+\sum_{m=1}^{M}i[m]Z_m[k],
\label{eq:dft_eq}
\end{equation}
\noindent where
\begin{equation}
Z_m[k]=Z[k]\text{exp}\left \{-j \frac{2\pi r \tau[m]}{K} \right \},
\label{eq:dft_eq2}
\end{equation}
\noindent and $i[m]$   for $m\in\{1,...,M\}$  is a binary variable, 
representing the WI of the $m^{th}$ mobile device with its unique  cyclic shift of $\tau[m]=mK_{cs}$, where  $ M\leq \left \lfloor{K/K_{cs}}\right \rfloor-1 $. Additionally, $Z_0$ is assumed to be always transmitted within the PDWCH group, helping  WRx  to retain   synchronization even if $i[m]=0,~ 	\forall m \in \{1,...,M\}$. In Eq. (\ref{eq:dft_eq2}) and in the rest of the article, modulo-$K$ indexing is assumed. The PDWCH signal structure   is illustrated in Fig. \ref{fig:structure}.

 In NM, the HPO is assumed to be kept ON during the  connected mode, therefore the implications of clock drift are likely to be small during the  w-cycles. In order to further reduce the impact of any possible residual clock drift,    the proposed receiver has two-stage mechanism to recover any potential symbol time offset (STO) and carrier frequency offset (CFO). Specifically, at the first stage, which is performed in pre-FFT domain (time domain),  maximum likelihood (ML)-estimator is applied to identify  STO and fractional CFO. After the synchronization of the  symbol timing, CP is removed correspondingly, and  the second stage is performed in the FFT domain. In this stage,   $N$-point  FFT ($N>K$)  is applied to jointly detect   integer CFO  and WI.  Moreover, because of decoding PDWCH in   every w-cycle,  WRx updates and compensates for    time and frequency offsets every w-cycle, hence sustaining orthogonality among the subcarriers. The involved processing aspects are described next, in details.

 \subsection{First Stage: Initial  Synchronization}
 The initial synchronization stage handles the potential STO and fractional CFO by exploiting  intrinsic redundancy
 in  CP of each OFDM symbol,  as originally proposed in  \cite{Beek1997}.  The CFO, normalized by the subcarrier spacing, is   decomposed into integer CFO component  ($\epsilon_i$)   and  the fractional CFO part  ($\epsilon_f$), where $\epsilon_i  \in \mathbb{Z}$ and  $-0.5 \leq \epsilon_f<0.5$. Similarly, STO ($\delta$) is the integer-valued relative time offset of the OFDM symbol in terms of sample instants.

\begin{figure}[t!]
\centering
\includegraphics[scale=3.5]{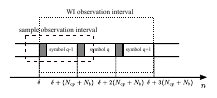}
\caption{Observation intervals for both synchronization and WI detection, assuming WI observation interval of three OFDM symbols, i.e., $x=3$. The gray parts of OFDM symbols are reflecting the CPs.}
\label{fig:window}
\end{figure}

WRx observes and collects samples of the  received signal ($r[n]$) within sliding sample observation window  with length of $N_{cp}+2N_{b}$. As shown in Fig. \ref{fig:window}, the window contains one OFDM symbol with unknown arrival time $\delta$. Then, based on  well-known  ML  technique \cite{Beek1997}, WRx estimates both $\epsilon_f$ and $\delta$.

The initial correction subsystem uses  a numerically controlled oscillator (NCO) to generate an appropriate complex conjugated phasor, which is then multiplying the received signal   to correct for the fractional CFO.  After fractional CFO compensation, $\hat{\delta}$ is applied to correct for positioning of the  CP removal. Due to the potential remaining timing uncertainty,  the truncated OFDM symbol contains either noise-corrupted copy of PDWCH data or   DL data payload. The block diagram of the first stage processing is illustrated in the left part of Fig. \ref{fig:wrxi}.

\subsection{Second Stage: Final Synchronization and Wake-Up Detection}
\label{section:joint}

  Assuming that the initial synchronization  provides   an adequate orthogonality between subcarriers, the output of an $N$-point FFT at the  $q^{th}$ OFDM symbol, considering only the $K$ used/active subcarriers, is represented as   
   $\textbf{R}_{q}=[R_{ q }{[1]},...,R_{q }[{K}]]^T$.  Under the assumption that $s$ is the index of the OFDM symbol carrying the PDWCH,   $R_{s}[k]$   can be written as 
\begin{equation}
R_{s}[k]=Y{[k-\epsilon_i]} H{[k-\epsilon_i]}\text{exp}\left \{-j\frac{2\pi\upsilon(k-\epsilon_i)}{N}\right \}+W[k],
\label{eq:received_signal_eq}
\end{equation}
 \noindent  where    $\upsilon$ denotes residual timing error, normalized by the sampling period, $H[k]$ is the channel frequency response at the $k^{th}$ subcarrier,   $\epsilon_i$ refers to the integer CFO, and $W[k]$ is  a circularly-symmetric white Gaussian noise process with average power $\sigma_{w}^2$.   Without loss of generality and for notational simplicity, we assume that $\upsilon$  is incorporated  into $H[k]$, therefore in the following,  $\text{exp}\left \{-j\frac{2\pi\upsilon k}{N}\right \}$ from Eq. ($\ref{eq:received_signal_eq}$) is removed and  absorbed as part of $H[k]$. Due to the ambiguity of the arrival time of  the   OFDM symbol carrying the PDWCH, the WRx must observe     a consecutive set of $x$    FFT outputs $\mathcal{R} =[\textbf{R}_{1},...,\textbf{R}_{x}]$, in order to locate the PDWCH symbol. The corresponding WI observation window  is  shown  in Fig. \ref{fig:window}.

In general, in the second stage, the WRx needs to perform three tasks, namely
1) to acquire the position of the PDWCH symbol ($s$),     
2) to estimate the integer CFO ($\epsilon_i$), and   
3) to decode the WI  ($i[m]$). 
In the following, we first introduce an estimation method  of the unknown parameters $(\hat s, \hat \epsilon_i)$ building on correlation-based processing and energy maximization. Then, we address how the actual detection of the WI bit ($i[m]$) can be carried out by utilizing the same correlation and energy quantities. The overall block diagram of the WRx processing, including both the first and the second stages, is shown in  Fig. \ref{fig:wrxi}.
\newcommand{\abs}[1]{\left|#1\right|}

 \subsubsection{Final Synchronization}
In the proposed approach,  the WRx    benefits   from ideal cyclic auto- and cross-correlation   properties of the ZC sequences by computing the received signal's power delay profile (PDP) through a frequency-domain matched filter. For this purpose, the squared absolute value of the  correlation of the received $q^{th}$ OFDM symbol with cyclic- and  and phase-shifted   root ZC sequences is given by
\begin{equation}\label{eq:pdp_energy_equation}
\Psi_q(l, \epsilon)=\abs{\sum_{n=0}^{K-1} r_{q}[n]z^*[{n+l}]\text{exp}\left \{j \frac{2\pi \epsilon}{K}(n+l) \right \} }^2,
\end{equation}
\noindent where $r_{q}[n]$ is the time-domain sequence corresponding to $R_q[k]$ and  $\epsilon$ represents the integer frequency shift of ZC sequence. Building on the ideal cyclic auto- and cross-correlation   properties of ZC sequences,   the $\Psi_q (l, \epsilon)$  for all   lags ($l$) reaches the maximum at the correctly estimated CFO ($\epsilon=\epsilon_i$) and PDWCH symbol index ($q=s$).  In order to reduce the implementation complexity, a hybrid time/frequency  domain method is considered for the second stage. Using the properties of FFT, $\Psi_q (l, \epsilon)$  can be equivalently computed as 
\begin{equation}
\Psi_q (l, \epsilon)=\abs{ \psi_q(l, \epsilon)}^2=\abs{\text{IFFT}\left \{R_q[k] Z^*[k-\epsilon]\right \}_l  }^2, 
\end{equation}
where $\psi_q(l, \epsilon)$   is the discrete periodic correlation function   of the received signal and the complex conjugate of frequency-shifted  version of the root ZC sequence, evaluated at lag $l$.

\begin{figure}[t!]
\centering
\includegraphics[scale=.89]{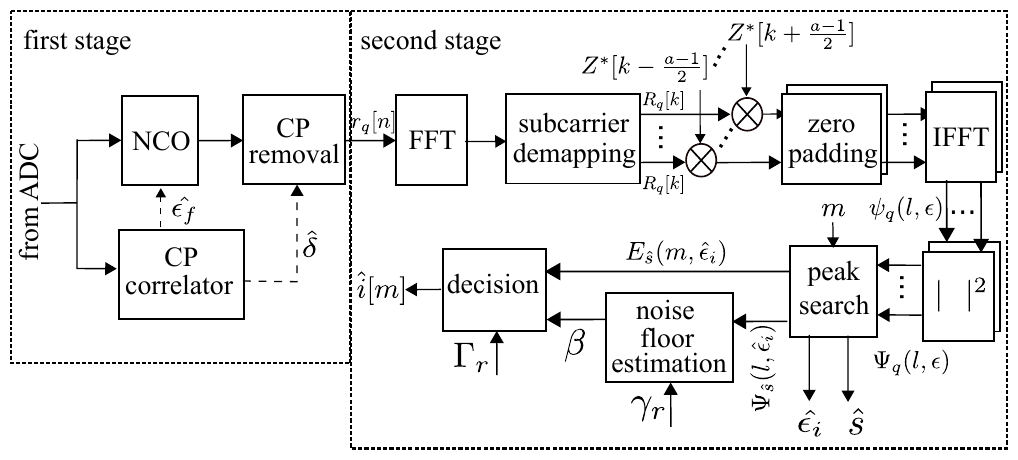}
\caption{Block diagram of the main processing components of the WRx;   ${\Gamma}_r$ and ${\Upsilon}_r$ are  pre-computed coefficients and stored in memory.}
\label{fig:wrxi}
\end{figure}

To further substantiate on the processing details, after removing   CP at the first stage, and obtaining initial synchronization,  $r_q[n]$ is translated to the frequency domain by using an $N$-point FFT, as shown in Fig. \ref{fig:wrxi}. The $K$ subcarriers corresponding to PDWCH are extracted from the output of  the FFT by using a subcarrier demapper. The result of subcarrier demapping ($R_q[k]$) is multiplied by the   complex-conjugated root ZC   sequences  with potential frequency offsets ($Z^*[k-\epsilon]$), and    the result  is properly oversampled by factor of $L$ by padding zeros in order to balance between detection performance and implementation complexity. Next, the IFFT block  transforms the product of  $R_q[k]$ and $ Z^*[k-\epsilon]$  from frequency into time domain. After that, the PDP samples are calculated by   squaring   the absolute value of the time-domain samples ($\psi_q(l, \epsilon)$).   The received  energy within the sliding window corresponding to $m^{th}$ interval $m\in \{0,...,M\}$, belonging to $q^{th}$ OFDM symbol   can therefore be written as
\begin{equation}
 E_q(m, \epsilon)=\sum_{l=mLK_{cs}}^{(m+1)LK_{cs}-1}\Psi_q (l, \epsilon). 
\end{equation}
Once the PDP samples of all potential frequency offsets of the root ZC sequence are obtained, all  $ E_q(0, \epsilon)$ per frequency offset per OFDM symbol are calculated, and the corresponding frequency offset and the index of the OFDM symbol of the maximum received energy   are chosen as estimates of $\epsilon_i$  and $s$, respectively.

\subsubsection{WI Detection}
Next we address how the above same energy quantities can be directly utilized in the actual WI detection. Under the assumption that the $LK_{cs}$ samples in the sliding window in the absence of WI  are uncorrelated  Gaussian noise samples with variance ${\sigma_{w}^2}$, the samples of  $\psi(l, \epsilon)$ also present Gaussian distribution with zero mean and  variance   of ${K}{\sigma_{w}^2}$. Consequently,  $\Psi (l, \epsilon)$ has a central Chi-squared distribution with $2$ degrees of freedom with  noise floor of $\beta={K}{\sigma_{w}^2}$. Therefore, the absolute WI detection threshold ($\Gamma$) can be calculated under the hypothesis of absence of WI as $P_{fa}=1-{F}_{1}(\Gamma)$ 
\noindent  where ${F}_{1}$ is  the cumulative distribution function (CDF) of $\Gamma$, and can be modeled as   a central Chi-squared random  variable with $2LK_{cs}$ degrees of freedom  \cite{proakis}. \noindent Without loss of generality,  we can assume that $\Gamma=\beta {\Gamma}_r $, where ${\Gamma}_r $ is the threshold relative to the noise floor $\beta$. By doing such, dependency of   ${F}_{1}({\Gamma}_r)$  on the noise variance is removed (i.e. $P_{fa}=1-{F}_{1}(\Gamma_r)$), and can be   expressed as  \DeclareRobustCommand{\rchi}{{\mathpalette\irchi\relax}}
\newcommand{\irchi}[2]{\raisebox{\depth}{$#1\chi$}}
 \begin{equation}
{F}_{1}({\Gamma}_r)= 1- \text{exp}\{-{\Gamma}_r\} \sum_{k=0}^{LK_{cs}-1}\frac{1}{{k\,!}}{{\Gamma}_r}^k,
\end{equation}
\noindent where  ${\Gamma}_r$  is a pre-computed coefficient which can thus be stored in memory.

Similarly, the noise power samples at the input to noise floor estimation follows a central Chi-square distribution with $2$ degrees of freedom,  expressed as
 \begin{equation}
{F}_{2}(\Upsilon_r)= 1- \text{exp}\{-\Upsilon_r\}.
\end{equation}
\noindent In above, $\Upsilon_r$ is the relative detection threshold   for noise floor estimation, and  is set as follows
\begin{equation}
P_{fa}=1-{F}_{2}(\Upsilon_r), 
\end{equation}
\noindent while the absolute noise floor threshold ($	\Upsilon$) can be computed    as
  \begin{equation}
\Upsilon=\frac{\Upsilon_r}{N}\sum_{l=1}^{N}\Psi_{\hat s} (l, \hat\epsilon_i),
\end{equation}
Finally $\beta$ can be estimated as \cite{theorylte}
  \begin{equation}
\beta=\frac{1}{N_s}{\sum_{\Psi_{\hat s} (l, \hat\epsilon_i)<\Upsilon}\Psi_{\hat s} (l, \hat \epsilon_i)},
\end{equation}
where the accumulation is over all samples less than $\Upsilon$, and $N_s$ is the number of available samples.

Then by utilizing a sliding window,  if the received energy  of the estimated OFDM symbol in $m^{th}$ interval ($ E_{\hat s}(m, \hat \epsilon_i)$)  exceeds the  WI detection threshold  ($\Gamma$), the $m^{th}$ WRx decodes $\hat i[m]=1$, otherwise $\hat i[m]=0$.  

To sum up, $(\hat s,\hat \epsilon_i)$ are obtained as follows
\begin{equation}
(\hat s,\hat \epsilon_i)=\underset{(q,\epsilon)\in 	\Theta}{\operatorname{arg \hspace{1mm} max}}\{E_q(0, \epsilon))\} 
\label{eq:hypothesis_equation}
\end{equation}
\noindent and then
\begin{equation}
\hat{i}[m]= \left\{ \,
\begin{IEEEeqnarraybox}[][c]{l?s}
\IEEEstrut
0,&$\text{for}~~ E_{\hat s}(m, \hat\epsilon_i)< \Gamma$ \\
1,&$\text{for}~~ E_{\hat s}(m, \hat\epsilon_i)\geq \Gamma$
\IEEEstrut
\end{IEEEeqnarraybox}
\right.
\end{equation}

\noindent  assuming that $(q,\epsilon)$ is restricted to a given parameter space $\Theta$. As can be seen, accurate realization of the estimator in Eq. (\ref{eq:hypothesis_equation}) requires a search over $x\times a $ values where $x$ denotes the number of considered FFT outputs and $a$ refers to the maximum span of subcarriers for which the integer CFO estimate is sought. For example, assuming that the integer CFO can be a maximum of $\pm2$ subcarriers, then $a=5$. The essential processing ingredients are summarized in Fig. \ref{fig:wrxi}.

\section{Power Consumption and Delay Analysis}
\label{Markov}
In this Section, the power states of NM are modeled as a semi-Markov process, and then its average buffering delay and  {average power consumption} are calculated. Furthermore, the mathematical models shown in this section give an insight into the operation and the parameter configuration  of the wake-up scheme. Our approach for analysing both the power consumption and delay characteristics of the wake-up scheme is similar to the DRX analysis presented in \cite{DRX_etsi}. Thus, our models and approach deliberately build on similar notations to the work in \cite{DRX_etsi},  for readers' convenience. {It is also noted that in the proposed WRx concept, the inevitable misdetections and false alarms increase  the  buffering delay and energy consumption, respectively, which are properly reflected in the analysis.}

 For analytically tractability,  the ETSI traffic model \cite{etsi} is applied, which is widely used in various analytical and simulation studies related to 3GPP mobile radio networks. In the employed traffic model, a packet service session contains one or several packet calls with exponentially distributed session inter-arrival time ($t_{s}$), while each packet call's arrival time ($t_{pc}$) follows exponential distribution. Moreover,  each packet call  consists of a sequence of packets with exponentially distributed packet inter-arrival time ($t_{p}$) within the packet call.

Additionally, based on the ETSI traffic model, at any time, the upcoming packet call belongs to either  ongoing session or is the initial packet call of a new session  with probabilities of $P_{os}$ and $P_{ns}$,  respectively. It is obvious that the number of packet calls per a session ($\eta_s$) and number of packets   per a  packet call  ($\eta_{pc}$) follow geometric distribution,    hence  $P_{ns}=1/	\mathbb{E}[\eta_s]$ and $P_{os}=1-P_{ns}$ \cite{DRX_etsi}.   In the rest of this article, the following equation   is used for statistical characterization of both the packet call and the packet service session
  \begin{equation}
\Pr[t_x \leq T ]=1-\Pr[t_x> T ]=1-e^{-\lambda_{x} T},
\end{equation}
where $\lambda_{x}$ represents either $\lambda_{pc}$ or  $\lambda_{s}$ which are  the mean packet call arrival rate and the mean session   arrival rate,  respectively. In addition, $t_x$ represents either $t_s$ or $t_{pc}$.

\subsection{Semi-Markov Model}
Four     power states, namely, active-on timer (S$_0$), active-inactivity timer (S$_1$), WRx-ON (S$_2$), and sleep (S$_3$)  are modeled by a semi-Markov process, as shown in Fig. \ref{fig:statess}.   Briefly, the states and corresponding transitions between them are described as follows:

\begin{figure}[t!]
\centering
\includegraphics[scale=1.6]{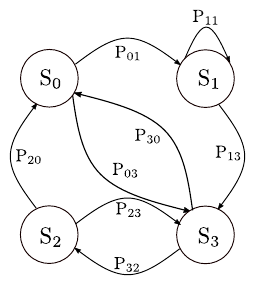}
\caption{Semi-Markov process for the state transitions of  NM. }
\label{fig:statess}
\end{figure}

\begin{itemize}
    \item 

\textbf{Active-on timer (S$_0$):}    when  NM is at  S$_0$, NM   is able to  receive data. Any packet arrival during  S$_0$  initiates   $T_I$, and NM moves to S$_1$. Otherwise at the expiry of $T_{ON}$, the NM moves to S$_3$;

\item 
\textbf{Active-inactivity timer (S$_1$):}   when NM is at S$_1$, $T_I$ is running, and if it is scheduled before the expiry of $T_I$, it restarts $T_I$ and remains at S$_1$, otherwise it transfers to S$_3$. Additionally, at S$_1$, NM functions fully, and is able to decode PDCCH and PDSCH;

\item 
\textbf{WRx-ON (S$_2$):}  at this state, WRx monitors PDWCH; if WRx receives WI=$0$,  NM transfers  to    S$_3$, otherwise  (WI=$1$)  it transfers to  S$_0$;

\item 
\textbf{Sleep (S$_3$):} at this state, NM  is asleep and is unable to receive any data.  Moreover,  NM at the end of sleep cycle moves  to S$_2$, unless  WRx has been at  S$_2$ and S$_3$ for $N_w$  consecutive  w-cycles, hence moving to S$_0$.
 \end{itemize}

 Because of the relativity short inter-packet arrival time within a packet call of eMBB-like sessions, for simplicity of analytical formulations, we can assume  that once a packet call starts,  NM serves all packets without transitioning to a new state. Similar assumption was adopted in DRX context in \cite{DRX_etsi}. Assume now that ${\text{P}_{kl}}$ denotes the transition  probability from ${\text{S}_{k}}$ to ${\text{S}_{l}}$ $\forall k,l \in\{0,1,2,3\}$.   When  NM  is at ${\text{S}_{0}}$, and if NM receives data before the expiry of $T_{ON}$, it moves to ${\text{S}_{1}}$, otherwise it moves to ${\text{S}_{3}}$. Thus, ${\text{P}_{01}}$ and ${\text{P}_{03}}$ can be calculated as 
  \begin{equation}
{\text{P}_{01}}=\Pr[t_{pc}\leq T_{ON}]P_{os}+\Pr[t_{s}\leq T_{ON}]P_{ns},
\end{equation}
\noindent and
 \begin{equation}
{\text{P}_{03}}=1-{\text{P}_{01}}. 
\end{equation}
 
 When  NM  is at ${\text{S}_{1}}$, it restarts $T_I$ if next packet starts before expiry of  $T_I$, otherwise it moves to ${\text{S}_{3}}$. Therefore,  ${\text{P}_{11}}$ and ${\text{P}_{13}}$ can be calculated as follows
 \begin{equation}
{\text{P}_{11}}=\Pr[t_{pc}\leq T_I]P_{os}+\Pr[t_{s}\leq T_I]P_{ns},
\end{equation}
\noindent and
 \begin{equation}
{\text{P}_{13}}=1-{\text{P}_{11}}. 
\end{equation}
 
  When   NM is at  ${\text{S}_{2}}$, it moves to ${\text{S}_{0}}$ either because of    false alarm or correct detection, otherwise it moves to ${\text{S}_{3}}$.     Therefore  ${\text{P}_{20}}$ and ${\text{P}_{23}}$ can be calculated as follows
  \begin{equation}
    \begin{split}
{\text{P}_{20}}=\Big(\Pr[t_{pc}> t_{sl}]P_{os}+\Pr[t_{s}> t_{sl}]P_{ns}\Big)P_{fa}\\
+\Big(\Pr[t_{pc}\leq t_{sl}]P_{os}+\Pr[t_{s}\leq t_{sl}]P_{ns}\Big)(1-P_{md}),
\end{split}
\end{equation}
\noindent and
\begin{equation}
{\text{P}_{23}}=1-{\text{P}_{20}},
\end{equation}
 \noindent where $t_{sl}$ is sleep period per w-cycle, i.e.,  $t_{sl}=t_c-t_{on}$.
 
 {Similarly,   ${\text{P}_{30}}=G({N_w})$ and ${\text{P}_{32}}=1-{\text{P}_{30}}$,  \noindent where $G({u})$, for   $u \in \{1,..., N_w\}$, is the probability of decoding WI for $u$ times as $0$, and can be calculated as follows}
    \begin{equation}
    \label{gu}
       \begin{split}
G({u})=\Bigg(\Big(\Pr[t_{pc}>t_{sl}](1-P_{fa})+\Pr[t_{pc}\leq t_{sl}]P_{md}\Big)P_{os}\\+\Big(\Pr[t_{s}>t_{sl}](1-P_{fa})+\Pr[t_{s}\leq t_{sl}]P_{md}\Big)P_{ns}\Bigg)^{u}.
\end{split}
\end{equation} 

Let $\text{P}_k$, $\forall k  \in\{0,1,2,3\}$ be the steady state probability that   NM is at     ${\text{S}_{k}}$. By utilizing the set of balance equations ($\text{P}_{k}=\text{P}_{l}\sum_{l=0}^{3}\text{P}_{lk}$) and the sum of probabilities equation  ($1=\sum_{k=0}^{3}\text{P}_{k}$), $\text{P}_k$, $\forall k  \in\{0,1,2,3\}$ can be solved easily. Due to space limitations, only the final equations are written herein, without intermediate steps, as 
\begin{equation}
\text{P}_{0}=\text{P}_{3}(\text{P}_{32}\text{P}_{20}+\text{P}_{30}),
\end{equation}
\begin{equation}
\text{P}_{1}=\frac{\text{P}_{3}\text{P}_{01}(\text{P}_{32}\text{P}_{20}+\text{P}_{30})}{1-\text{P}_{11}},
\end{equation}
\begin{equation}
\text{P}_{2}=\text{P}_{3}\text{P}_{32},
\end{equation}
\begin{equation}
\text{P}_{3}=\frac{1-\text{P}_{11}}{(\text{P}_{32}\text{P}_{20}+\text{P}_{30})(1+\text{P}_{01}-\text{P}_{11})+(1-\text{P}_{11})(1+\text{P}_{32})}.
\end{equation}

The    holding times for ${\text{S}_{k}}$ are represented as  $ \omega_k $    $\forall k \in\{0,1,2,3\}$. For ${\text{S}_{0}}$, NM may stay for $T_{ON}$ at this state (no arriving packet), or due to a packet arrival before expiry of $ T_{ON} $, it moves to ${\text{S}_{1}}$. Therefore, $ \omega_0(t) $ can be written as a function of time as follows  
  \begin{equation}
 \omega_0(t) = \left\{ \,
\begin{IEEEeqnarraybox}[][c]{l?s}
\IEEEstrut
t,&$\text{for}~~ t\leq T_{ON} $ \\
T_{ON},&$\text{for}~~ t>T_{ON} $
\IEEEstrut
\end{IEEEeqnarraybox}
\right.
\end{equation}
and thus 
     \begin{equation}
  \begin{split}
\mathbb{E}[\omega_0]=  P_{os}\int_{0}^{\infty}\omega_0(t)    f_{pc}(t)dt+P_{ns}\int_{0}^{\infty}\omega_0(t)    f_{s}(t)dt \\
=P_{os}\frac{(1-e^{-\lambda_{pc}T_{ON} })}{\lambda_{pc}}+P_{ns}\frac{(1-e^{-\lambda_{s}T_{ON} })}{\lambda_{s}},
 \end{split}
\end{equation}
  \noindent where $ f_{pc}(t)$ and $ f_{s}(t)$ are the probability density functions   of   exponential distributions of inter-packet call  and inter-session arrival time, respectively, which are of the form    $ f_{pc}(t)=\lambda_{pc}e^{-\lambda_{pc}t}$ and $ f_{s}(t)=\lambda_{s}e^{-\lambda_{s}t}$.
  
  Furthermore, with the assumption   of utilizing    stop-and-wait hybrid automatic request flow control algorithm (which can be modeled as an M/M/$8$  queuing model), $ \mathbb{E}[\omega_1]$  can be written as follows \cite{Kleinrock}  
   \begin{equation}
\mathbb{E}[\omega_1]=  \mathbb{E}[t_{pr}]+\mathbb{E}[{t_i}], 
\end{equation}
 \noindent where $ \mathbb{E}[t_{pr}]$ is the average processing time for transmitting all packets of a packet call, and $\mathbb{E}[{t_i}]$ is the average time of inactivity timer  after serving an ongoing packet call, referred to  as the average inactivity period. The processing time   consists of serving 
$N_{p}$ packets with a  per-packet service time of  $t_{ser}$, which is the  time interval from transmission of a packet from gNB  to getting ACK from the mobile device. According to  Wald's Theorem \cite{Nelson} 
    \begin{equation}
\mathbb{E}[t_{pr}]=\mathbb{E}[N_p]\mathbb{E}[t_{ser}]=\frac{\eta_{pc}}{\lambda_p},
\end{equation}
 \noindent where $\eta_{pc}$ and $\lambda_p$ are the average number of packets within  a packet call and the mean packet arrival rate within a packet call,  respectively.  Additionally, if   a packet arrives before $T_I$, inactivity period equals the  inter-packet call time, otherwise  the inactivity period equals $T_I$. Therefore, the inactivity period  ($t_i$) can be calculated as a function of $t$ as follows
\begin{equation}
t_i(t)= \left\{ \,
\begin{IEEEeqnarraybox}[][c]{l?s}
\IEEEstrut
t,&$\text{for}~~ t\leq T_I$ \\
T_I,&$\text{for}~~ t>T_I $
\IEEEstrut
\end{IEEEeqnarraybox}
\right.
\end{equation}

\noindent while $\mathbb{E}[t_i]$ can be expressed as

   \begin{equation}
  \begin{split}
\mathbb{E}[t_i]=  P_{os}\int_{0}^{\infty}t_i(t)    f_{pc}(t)dt+P_{ns}\int_{0}^{\infty}t_i(t)    f_{s}(t)dt \\
=P_{os}\frac{(1-e^{-\lambda_{pc}T_I })}{\lambda_{pc}}+P_{ns}\frac{(1-e^{-\lambda_{s}T_I })}{\lambda_{s}}.
 \end{split}
\end{equation}

 \noindent Additionally, the holding times for ${\text{S}_{1}}$ and ${\text{S}_{2}}$ are fixed, and therefore  $\mathbb{E}[\omega_2]= t_{on} $ and $\mathbb{E}[\omega_3]=t_{sl}$.    
  
  \subsection{Average  Power Consumption Calculation}

 In order to develop an analytical  model for power saving achieved  by   NM, we assess the  average power consumption of NM ($\mathbb{E}[\text{PW}]$). Assuming   $\text{PW}_{k}$, $ k  \in\{0,1,2,3\}$, denotes the  power consumption of NM at state $\text{S}_{k}$, the average power consumption  can be expressed as
\begin{equation}
\mathbb{E}[\text{PW}]=\frac{\text{E}}{\text{T}},
\end{equation}
\noindent where  $\text{E}$ denotes the consumed average energy during overall observation time ($\text{T}$), and can be calculated as
   \begin{equation}
\text{E}=e_t+\sum_{n=0}^{3}\text{P}_{n}\mathbb{E}[\omega_n]\text{PW}_{n}.
 \end{equation}
 \noindent In above, $e_t$ is the average required energy for transitional states, and can be calculated as $e_t=\text{P}_{2}\text{P}_{20}(\text{PW}_{3}(t_{of}-t_{su})+e_{su})+\text{P}_{3}\text{P}_{30}e_{su}+(\text{P}_{1}\text{P}_{13}+\text{P}_{0}\text{P}_{03})e_{pd}$, where $e_{su}$ and $e_{pd}$ are the required energies for transitioning of NM from sleep to fully active and vice versa, respectively. 
 
Additionally, $T$  can be calculated as follows
  \begin{equation}
 \text{T}=t_t+\sum_{n=0}^{3}\text{P}_{n}\mathbb{E}[\omega_n],
 \end{equation}
 \noindent where  $t_t$ is the mean overall time period that NM spends on transitional periods, i.e., $t_t=\text{P}_{2}\text{P}_{20}t_{of}+\text{P}_{3}\text{P}_{30}t_{su}+(\text{P}_{1}\text{P}_{13}+\text{P}_{0}\text{P}_{03})t_{pd}$.

 In next sections, concrete example numerical values of  power  and energy consumption   of different states  are presented (cf. particularly Table \ref{pc}).

   \subsection{Average  Delay Calculation}
  We assume that the  packets that arrive during  ${\text{S}_{2}}$ and  ${\text{S}_{3}}$  are  buffered at gNB until   NM enters to ${\text{S}_{0}}$, causing some buffering delay.  {The overall delay experienced by end users consists of server delay, core network delay, buffering delay and scheduling delay. However, considering all these in analytical work complicates the system model largely making it intractable. Thus, for the purpose of the analytical work, the server and core network delays are ignored. Furthermore,  without loss of generality,  we assume that the radio access network experiences  non-fully-loaded  traffic conditions, and thus all packets that arrive  during ${\text{S}_{0}}$ and  ${\text{S}_{1}}$  can be served promptly without further scheduling delay.  Ignoring the packet scheduling contribution  to  the  overall delay is  a  valid  assumption  when  there  are  adequate  radio resources per TTI that can accommodate the data packets of target UEs, while in other cases it is a simplification.}

  Now, in order to have a feasible and intuitive delay analysis and expressions, a packet arrival time during ${\text{S}_{2}}$ and  ${\text{S}_{3}}$ is assumed to belong to one of the two following intervals,  1) between first and ${N}_{w}$ w-cycles, or 2) during ${t}_{of}$ after  ${N}_{w}$ w-cycles, where for all cycles, WI=$0$ is decoded. In the former case, the average buffering delay (${d}_{1}(u)$) is caused by a packet that arrives on ${u}^{th}$ w-cycle, where $u \in \{1,..., N_w\}$, with the numbering starting from the first w-cycle    straight after transition from S$_0$ or S$_1$  to S$_3$. This delay can be calculated as follows 
\begin{equation}
  \begin{split}
{d}_{1}(u)=    \sum_{n=1}^{N_w-u+1}(1-P_{md}) { P_{md} }^{ n-1 }\int_{0}^{{t_{sl}}}(nt_c+t_{of}-t)\\
 \Big(P_{os}f_{pc}(t)+P_{ns}f_{s}(t)\Big)dt+
 { P_{md} }^{(N_w-u+1)}\int_{0}^{{t_{sl}}}\\\Big({(N_w-u+1)}t_c+t_{of}-t\Big)\Big(P_{os}f_{pc}(t)+P_{ns}f_{s}(t)\Big)dt.
 \end{split}
\label{eq:d1_equation}
\end{equation}
 While in the latter case,  the average buffering delay of a packet (${d}_{2}$) arriving after  ${N}_{w}$ w-cycles, where for all w-cycles, WI is decoded as $0$, can be calculated as 
   \begin{equation}
  \begin{split}
{d}_{2}=  
\int_{0}^{t_{of}}(t_{of}-t)\Big(P_{os}f_{pc}(t) +
P_{ns}  f_{ns}(t)\Big)dt.
 \end{split}
\label{eq:d2_equation}
\end{equation}

Finally, by combining Eq. (\ref{eq:d1_equation}) and  Eq.   (\ref{eq:d2_equation}), and using the corresponding probabilities of first and second intervals, the overall average  buffering delay ($\mathbb{E}[\text{D}]$), imposed by the wake-up scheme, can be written as 
   \begin{equation}
   \label{eq:en_}
\mathbb{E}[\text{D}]= ({\text{P}_{2}}+{\text{P}_{3}})\bigg(
\sum_{u=1}^{N_w} G(u-1){d}_{1}(u)+G(N_w){d}_{2}  \bigg).
\end{equation}
Due to resulting fairly long expressions, the explicit substitutions of     (\ref{gu}),  (\ref{eq:d1_equation}) and (\ref{eq:d2_equation}) into    (\ref{eq:en_}) are omitted.

\section{Overall Receiver Architecture and \\Implementation Aspects}
\label{h1}
For the completeness of the study, before going into the numerical performance evaluations, we provide in this section selected implementation insight regarding how the proposed WRx processing can be efficiently incorporated into the overall new modem (NM) hardware. Even though the finest implementation details and optimization are outside the scope of this paper, fairly elaborate implementation aspects are reported and discussed in order to understand the concept feasibility and particularly the realistic power consumption characteristics of the WRx unit. {Further specifics regarding the power consumption details are provided as supplementary material in the Appendix.}
\subsection{WRx Implementation Fundamentals}
\begin{figure}[t!]
 \centering
  \includegraphics[scale=1.03]
  {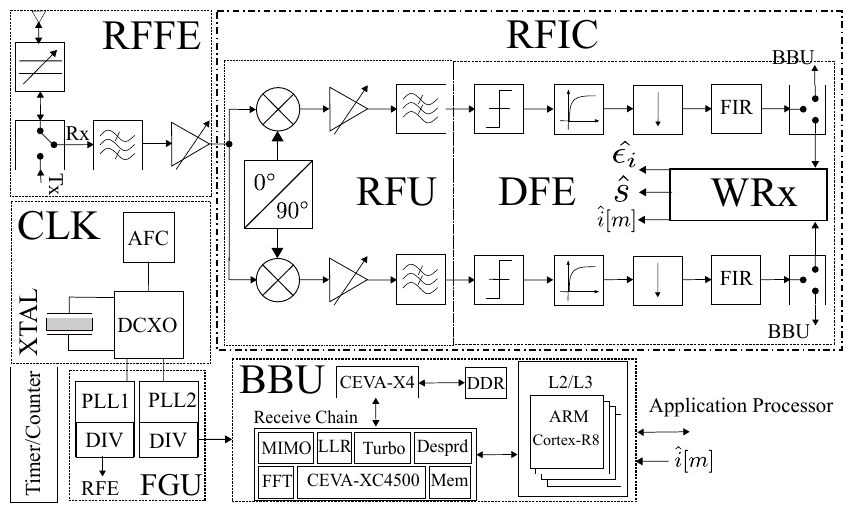}
 \caption{Main building blocks of the new modem (NM). Due to the versatility of the modern RFICs, the WRx processing is embodied directly in the NM RFIC.}
\label{fig:block_NM_schematic}
\end{figure}

The NM, illustrated at principal implementation level in Fig. \ref{fig:block_NM_schematic}, provides all corresponding radio and signal processing tasks for the wake-up concept, while naturally also handles the  typical cellular subsystem functionalities. In general, typical cellular BB  processing     needs dedicated digital {application-specific integrated circuits (ASICs)} as well as  one or multiple processors to perform the defined signal processing and protocol stack tasks. However, owing to the low processing complexity of the proposed wake-up scheme, the WRx functionalities can be implemented directly in {radio frequency integrated circuit (RFIC)}, providing reduced power consumption, higher integrity, and fast access as well as essentially negligent  start-up and power-down  periods. Furthermore, as fetching data from external processor through RF-digital interface is time and power consuming,   the RFIC-based implementation of the wake-up detection is favored also in this sense.

The embedment of the WRx on RFIC is generally enabled by the recent advances in the low-cost and power-efficient  CMOS  technology, allowing also selected digital functionalities to be implemented directly on RFIC. Additionally, modern RFICs are highly reconfigurable over most of  their building blocks, providing   high flexibility. Such flexibility is particularly interesting in the context of the wake-up scheme, as both the bandwidth and the gain  performance can be adapted for the reception of the wake-up signaling. Furthermore, the placement of the WRx at digital domain of the RFIC   permits multiple filter bandwidths to be included without any essential penalty in the silicon area. Thus, the WRx helps for maintaining synchronization of the overall NM, while carries the wake-up signal processing. Similar to the BBU,  WRx receives the complex I/Q samples from digital front-end (DFE) on chip. Furthermore, WRx utilizes the processing and memory capabilities of the RFIC, together with a dedicated FFT/IFFT block, in order to detect the WI. The configuration of the radio frequency front-end (RFFE) and RFIC is controlled at several stages, through applicable software, in WRx and BBU via control links.

{For further details related to task specific aspects, the readers are referred to the Appendix.}

\begin{table}[t!]
\small
\centering
\caption{Measured power consumption of NM subcomponents in different  states  (@$20$ MHz channel BW); all power values are in mW }
\newcolumntype{L}[1]{>{\raggedright\let\newline\\\arraybackslash\hspace{0pt}}m{#1}}
\newcolumntype{C}[1]{>{\centering\let\newline\\\arraybackslash\hspace{0pt}}m{#1}}
\newcolumntype{R}[1]{>{\raggedleft\let\newline\\\arraybackslash\hspace{0pt}}m{#1}}
\begin{tabular}{|L{.2cm}|L{.2cm}|L{.2cm}|L{.2cm}|L{.2cm}|L{.2cm}|L{.2cm}|L{.2cm}|L{.2cm}|L{.2cm}|L{.2cm}|}
\hline
\begin{turn}{+90}\textbf{power state}\end{turn}& \begin{turn}{+90}\textbf{RFFE}\end{turn}&\begin{turn}{+90}\textbf{RFU}\end{turn} &\begin{turn}{+90}\textbf{DFE}\end{turn}& \begin{turn}{+90}\textbf{WRx}\end{turn}  & \begin{turn}{+90}\textbf{CLK}\end{turn} & \begin{turn}{+90}\textbf{FGU }\end{turn} & \begin{turn}{+90}\textbf{BBU }\end{turn}& \begin{turn}{+90}\textbf{Interface}\end{turn}& \begin{turn}{+90}\textbf{Others}\end{turn}& \begin{turn}{+90}\textbf{PW}\end{turn}\\ \hline
 \multicolumn{1}{|c|}{S$_3$} & \multicolumn{1}{c|}{$0$} &$0$&0 &0&\multicolumn{1}{c|}{$10$}&\multicolumn{1}{c|}{$4$}&\multicolumn{1}{c|}{$0$}&\multicolumn{1}{c|}{$0$}&\multicolumn{1}{c|}{$2$}&\multicolumn{1}{c|}{$16$}\\\hline
  \multicolumn{1}{|c|}{S$_2$} & \multicolumn{1}{c|}{$6$} & \multicolumn{1}{c|}{$7$}&\multicolumn{1}{c|}{$10$} &\multicolumn{1}{c|}{$11$}&\multicolumn{1}{c|}{$10$}&\multicolumn{1}{c|}{$10$}&\multicolumn{1}{c|}{$0$}&\multicolumn{1}{c|}{$3$}&\multicolumn{1}{c|}{$1$}&\multicolumn{1}{c|}{$57$}\\\hline
   \multicolumn{1}{|c|}{S$_0$/S$_1$ } & \multicolumn{1}{c|}{$11$} &\multicolumn{1}{c|}{$53$}&\multicolumn{1}{c|}{$38$} &\multicolumn{1}{c|}{$0$}&\multicolumn{1}{c|}{$10$}&\multicolumn{1}{c|}{$24$}&\multicolumn{1}{c|}{$670$}&\multicolumn{1}{c|}{$27$}&\multicolumn{1}{c|}{$17$}&\multicolumn{1}{c|}{$850$}\\\hline
\end{tabular}\label{pc}
\end{table}

\subsection{Measurement-based Power Consumption}
 \label{h2}
Next, measurement based experimental power consumption model of the overall NM is presented, building on 16nm CMOS implementation when it comes to the RFIC. As defined in Section \ref{Markov}, the NM operation contains essentially four states, namely, S$_0$, S$_1$, S$_2$ and  S$_3$. In general, in a given state, the overall power consumption comprises of a number of components. Furthermore, depending on the specific NM state,  some components may be either ON   or OFF. In general, within S$_0$ or S$_1$, NM's power consumption vastly depends on the operating bandwidth, while the power consumption values in S$_2$ and S$_3$ are essentially independent of the bandwidth since a fixed resource of $128$ subcarriers each of width $15$~kHz are assumed for the PDWCH, {as further detailed in the Appendix}. 

Table \ref{pc} represents an experimental    power consumption breakdown of the NM'\textsc s main hardware components assuming the basic channel bandwidth of $20$ MHz.  Our  methodology  for profiling the energy consumption builds on the physical power measurements, assessing the current and voltage of the resistors on the power supply rails of the relevant components on real hardware. As can be observed through Table \ref{pc}, the proposed WRx, including both radio and digital processing, consumes  only some $\text{PW}_{2}=57$ mW of power. This is generally a very low power consumption figure. Additionally, Table \ref{pc} details how this figure is contributed by the different subcomponents or modules. {It is noted that the power consumption values utilized to evaluate the proposed scheme, reported in Table \ref{pc}, are based on actual prototype hardware, while certainly the finalized products may still differ, being further improved and optimized in their final hardware and software solutions. Hence, these numbers utilized in the paper can be considered indicative, and representative, while the final products are beyond our scope. Furthermore, for readers' convenience, some further clarifications and details regarding the different modules and their design choices, considered in the power consumption experiments are provided as supplementary material in the Appendix.}

Finally, it is noted that for the purpose of  synchronization,   during sleep period, RFFE, RFU and BBU    are switched off, but HPO is ON, and a timer/counter is used to track the location of the starting symbol of the PDWCH subframe. Therefore, the energy expenditure for sleep state in NM is slightly higher than that in a typical cellular subsystem. Power consumption of cellular subsystem is mainly dependent on its implementation.   The overall NM entity consumes about $e_{su}=4.6$ mJ and $e_{pd}=3.1$ mJ in average for the duration of the corresponding    start-up   and power-down times of $t_{su}=12$ ms and  $t_{pd}=8$ ms, respectively. {Moreover, as discussed in Section \ref{section:wus}, the proposed NM assists the BBU to obtain synchronization immediately, leading to   $t_{sync}\approx 0$ (see Fig. \ref{fig:realisticdrx}).}

\section{Numerical Results and Analysis}
 \label{section:simulation}
 Numerical evaluations are next  conducted to assess and illustrate the performance of the presented  wake-up scheme. The wake-up receiver detection performance is first analysed, through empirical simulations, followed by the system energy saving and latency evaluations where also comparisons to the analytical results are made.

\subsection{Basic Assumptions}
\label{basic}
 The radio frame and assumed hardware settings for simulations and numerical evaluations follow directly the assumptions made already in Sections \ref{section:wus}-\ref{h1}. Specifically, the basic  $20$~MHz carrier bandwidth case with $15$~kHz subcarrier spacing is considered, while the radio frame duration is assumed to be 10~ms. For channel modeling, we adopt the $3$GPP extended pedestrian A model (EPA) \cite{epa}.

Each simulation scenario lasts for $100,000$ frames, and is repeated for $10,000$ times in order to collect statistically reliable results.  Each PDWCH group  has $M=7$ users, utilizing ZC sequences with length of $K=117$ and root index of   $r=31$. Moreover, we assume that $K_{cs}=13$.

Stemming from the {high-precision} oscillator assumption, the oscillator deviation can be assumed to be within  $\pm 2.00$~ppm  in room temperature. Thus, the clock drifting and CFO are very small within the duration of a single w-cycle. Even at a higher carrier frequency of say $5.0$~GHz, the maximum CFO stemming from the $\pm 2.00$~ppm accuracy  is $\pm 20.0$~kHz, corresponding to the case that the gNB's LO and the  NM's LO both have the largest allowed yet opposite-sign frequency deviations. This is less than twice the assumed subcarrier spacing, thus we set the value of the configurable parameter $a$  in the integer CFO estimation to $a=5$. Additionally, $x=3$ OFDM symbols is considered.

In general, in the energy saving and latency evaluations, the WRx configuration in terms of the w-cycle and different timers as well as the traffic model assumptions have all an impact on achievable performance figures. Baseline parameter configurations and traffic model assumptions, following those in   \cite{DRX_etsi}, \cite{Waqas}, are listed in Table \ref{tab:parameters}. Many of these parameter values are also varied in the evaluations, in order to better understand and assess the power consumption and latency characteristics of the considered wake-up based system, and the potential involved dependencies. Finally, the measurement-based power consumption values, shown in Table \ref{pc} for the different states, are used.

\begin{table}[t!]
\scriptsize
\renewcommand{\arraystretch}{1.3}
\caption{Default evaluation parameters for power consumption and latency evaluations, following \cite{DRX_etsi}, \cite{Waqas}. Many of the parameters are also varied in the evaluations.}
\label{tab:parameters}
\centering
\begin{tabular}{|c|c|}
    \hline
   {Parameter}  & { Value}   \\
    \hline
    \hline
      wake-up cycle, $t_c$& $10$ ms  \\
    \hline
on-time duration,  $T_{ON}$& $1$ ms  \\
    \hline
 inactivity timer, $T_{I}$& $12$ ms  \\
    \hline
       offset time,  $t_{of}$& $15$ ms  \\
    \hline
average session inter-arrival time,    $1/\lambda_s$& $60$ s  \\
    \hline
  average packet-call  inter-arrival  time,         $1/\lambda_{pc}$& $0.2$ s  \\
    \hline
       average packet  inter-arrival  time,     $1/\lambda_{p}$& $0.01$ s  \\
    \hline
average number of packet calls per session,         $\eta_{s}$& $6$   \\
    \hline
average   number of packets per packet-call,      $\eta_{pc}$& $50$   \\
    \hline
\end{tabular}\vspace{-2mm}
\end{table}

\subsection{Wake-up Receiver Detection Performance}
\begin{figure}[t!]
\centering
     \includegraphics[width=0.49\textwidth]{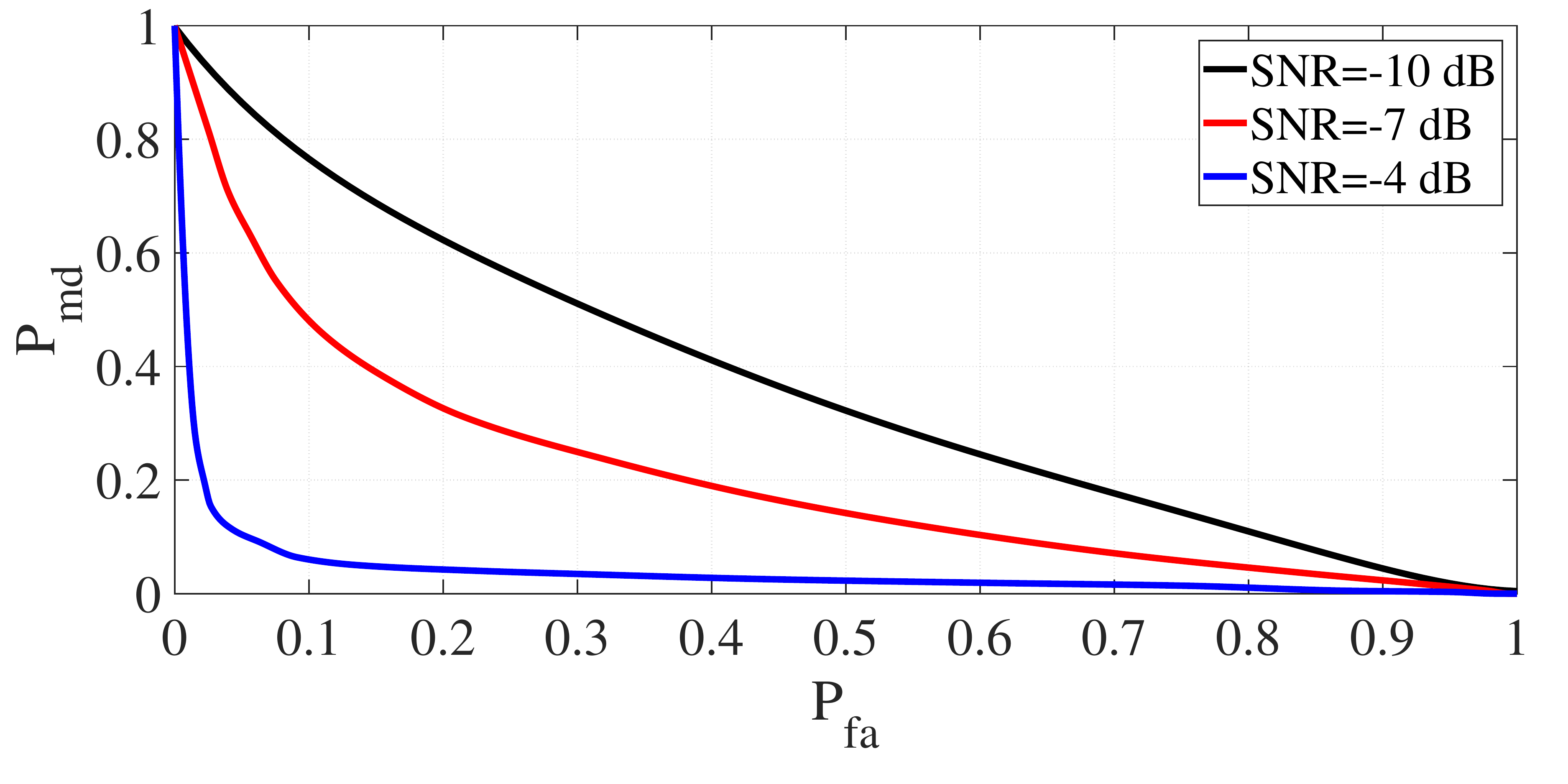}
      \caption{Wake-up signal detection related ROC curves for three different SNRs.}
       \label{roc}
\end{figure}

We begin by addressing the reliability of the wake-up signal detection at WRx, together with the achievable synchronization performance, with specific emphasis on challenging low-SNR conditions. Fig. \ref{roc} shows the receiver operating characteristic  (ROC) curve  of the proposed wake-up signaling and detection scheme under the EPA channel model at three different SNR values of $-10$~dB, $-7$~dB, and $-4$~dB, respectively.  As it can be expected, for all three ROC curves, there is a clear trade-off between the achievable probability of missed detection, $P_{md}$, and the probability of false alarm, $P_{fa}$. By comparing the ROC   curves corresponding to the different SNRs, and assuming that the target false alarm probability is $P_{fa}=10$\%, we can observe that very reliable wake-up signal detection can still be obtained at an SNR of $-4$~dB while the detection performance is then sharply degrading for lower SNRs of $-7$~dB and $-10$~dB.

\begin{figure}[t!]
\centering
     \includegraphics[width=0.49\textwidth]{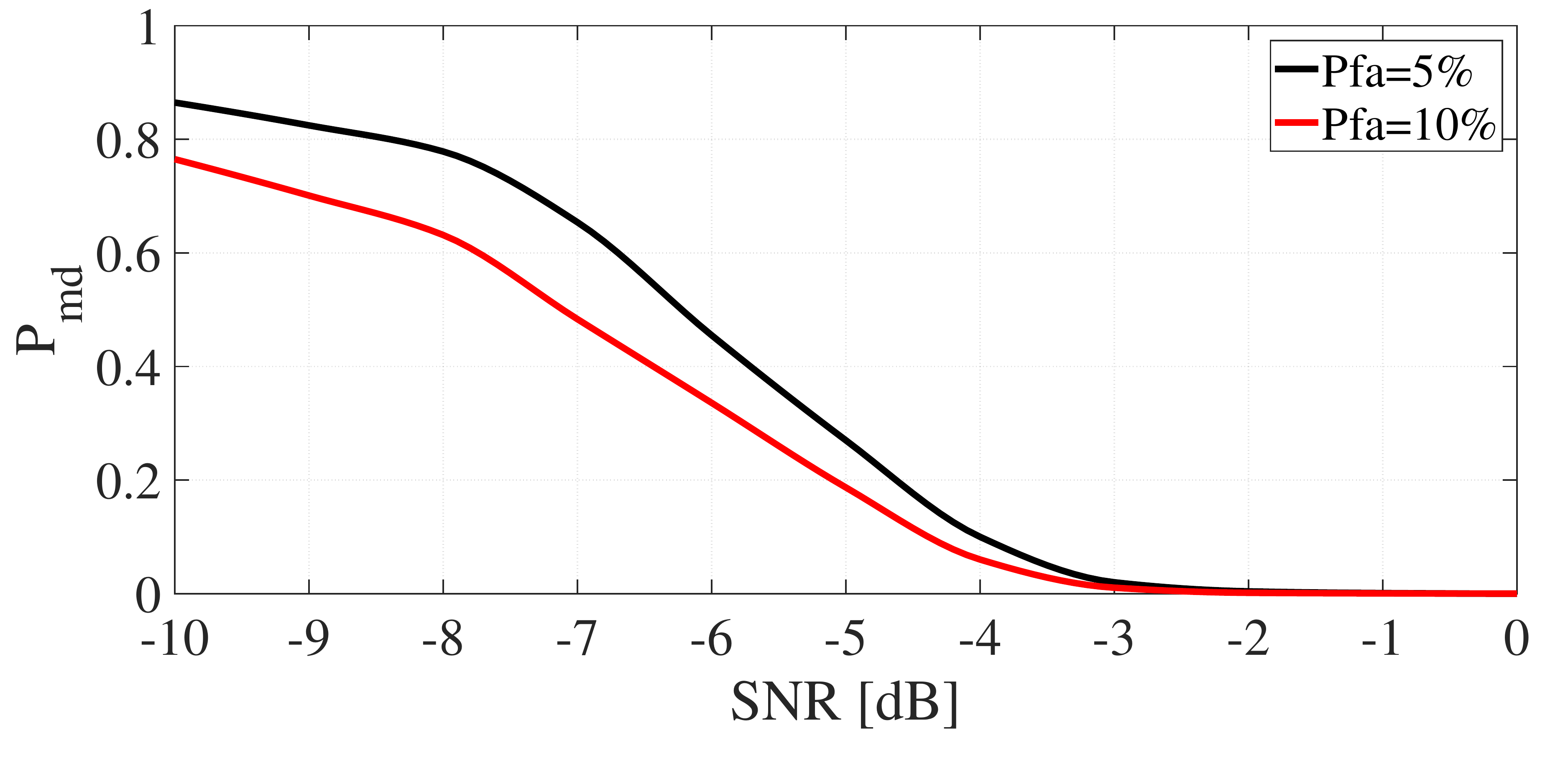}
      \caption{Misdetection rate as a function of SNR, for two different false alarm constraints.}
       \label{p_MD_SNR}
\end{figure}

The further assess the detection performance and its dependency on the SNR as well as on the $P_{fa}$ constraint, additional numerical results are provided in Fig. \ref{p_MD_SNR}. Specifically, the figure shows the behavior of the $P_{md}$ for varying SNR with two different $P_{fa}$ constraints, namely 5\% and 10\%. The figure clearly illustrates that very reliable wake-up signal detection can be achieved down to SNRs in the order of $-2$ $\dotsc$ $-5$~dB, depending on the exact detection probability target. For SNRs yet lower than these, the performance degrades clearly. Additionally, as expected, the misdetection rate for a lower false alarm rate constraint is higher than the corresponding misdetection rate of  higher false alarm rate, at any given SNR. This implies the need for configuring the target false alarm rate individually for different use cases based on their energy and delay requirements.

\begin{figure}[t!]
\centering
\minipage[t]{0.5\linewidth}
\begin{subfigure}[t]{1.0\linewidth}
    \includegraphics[width=\textwidth]{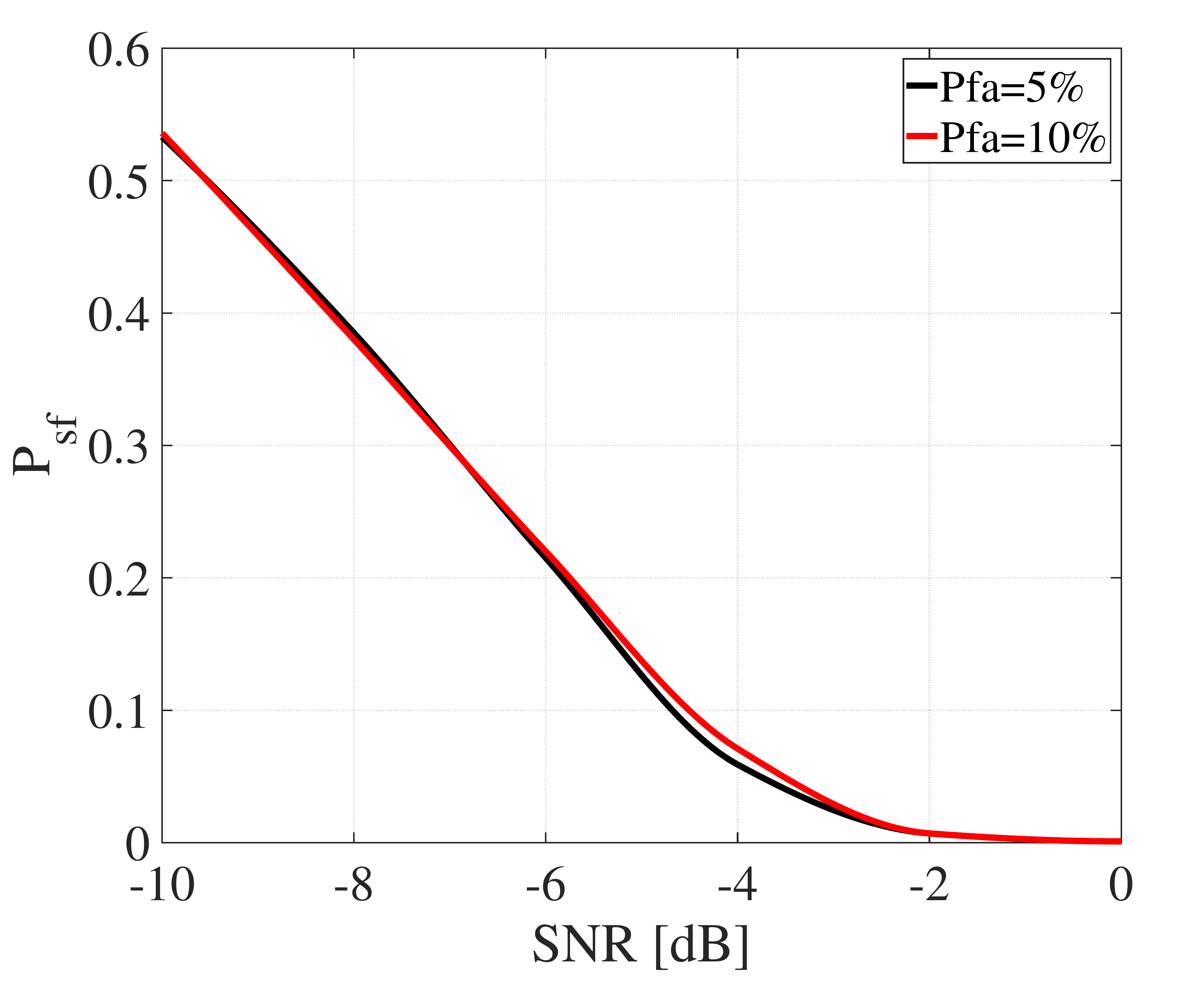}
  \caption{}
\end{subfigure}
\endminipage
\minipage[t]{0.5\linewidth}
\begin{subfigure}[t]{0.97\linewidth}
    \includegraphics[width=\linewidth]{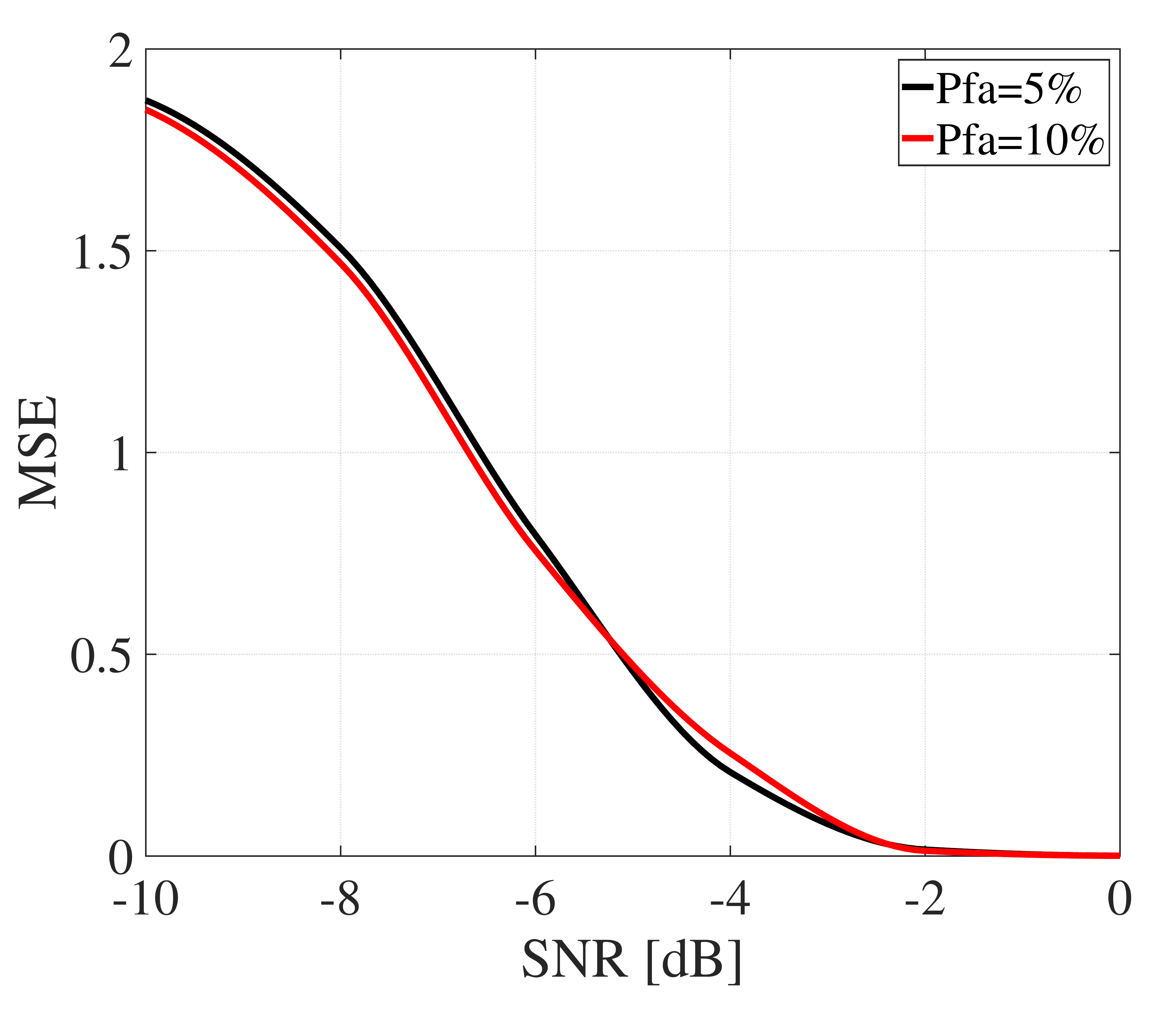}
  \caption{}
\end{subfigure}
\endminipage
\caption{Synchronization failure rate and MSE of $\hat{\epsilon}_i$, normalized by the subcarrier spacing, under the two different false alarm constraints.}\label{fig:MSE}
\end{figure}

Next, Fig. \ref{fig:MSE}(a) and  Fig. \ref{fig:MSE}(b) illustrate the obtained synchronization failure rate (${P}_{sf}$) performance, and the achieved mean squared error (MSE) of the CFO estimate as a function of SNR, respectively. As can be observed, the failure rate reduces for increasing SNR, while very high-quality synchronization can be obtained down to SNRs of some $-4$~dB. Interestingly, the synchronization failure rate is not dependent on $P_{fa}$ constraint. The main reason is that $\epsilon_i$ is estimated without utilizing the target $P_{fa}$ based threshold. We note that the detection results in Figs. \ref{roc} and  \ref{p_MD_SNR} contain also the impacts of the synchronization procedures, hence illustrating reliably the achievable detection performance.

Moreover, because of utilizing PDP values for selecting or identifying the PDWCH symbol index, the probability of detection of the correct PDWCH OFDM symbol is almost $100$\%, even for low SNRs in the order of -$10$ dB, {\color{black}and is thus not explicitly illustrated}. The PDPs of other OFDM symbols are extremely low compared to that of the PDWCH symbol. It is in general very vital to have large PDWCH detection rate, since misdetection of the PDWCH symbol index would directly increase both the false alarm rate and the misdetection rate of the wake-up signaling. We note that also the PDWCH symbol index detection is built in to the results in Figs. \ref{roc} and  \ref{p_MD_SNR}, while not illustrated separately. 

{It is additionally noted that the false alarm and misdetection rates are essentially independent of the w-cycle, unless the w-cycle is very long, say tens of seconds, which could cause the frequency and/or time offsets to be eventually larger than the chosen values of $x$ and $a$. In contrast, it is also noted that the overall numbers of false alarm and misdetection events, within a given overall time window, do indeed increase when the w-cycle reduces. Similarly, the impact of the value of $T_{ON}$ on the false alarm and misdetection rates is essentially negligible. This is due to the fact that even when identifying WI=1, the WRx is assumed to carry on the PDWCH reception, according to its w-cycle, until receiving the target PDCCH message.}

\subsection{Energy Saving and Latency Performance}

Next, the energy consumption and latency characteristics of the wake-up based system are assessed. We use and evaluate the analytical results, obtained in Section \ref{Markov}, while also simulate and compare corresponding empirical results. {The baseline values for the wake-up scheme parameters are those shown in Table \ref{tab:parameters}, while many of them are also again varied.} Additionally, the measurement based power consumption values, shown in Table \ref{pc} for the different states, are used. Regarding the assumed wake-up signal detection performance in the analysis, four pairs of false alarm and misdetection rates are considered, building on the results obtained in the previous subsection. Specifically, we consider the following examples cases, obtained from Fig. \ref{p_MD_SNR}:
\begin{itemize}
    \item $P_{fa}=5$\% \hspace{2.65mm} and \hspace{1mm} $P_{md}=1$\% \hspace{1mm} (SNR of $-2.6$~dB),
 \item $P_{fa}=10$\% \hspace{1mm} and \hspace{1mm} $P_{md}=1$\% \hspace{1mm} (SNR of $-3$~dB),
 \item $P_{fa}=10$\% \hspace{1mm} and \hspace{1mm} $P_{md}=5$\% \hspace{1mm} (SNR of $-3.8$~dB),
 \item $P_{fa}=5$\% \hspace{2.65mm} and \hspace{1mm} $P_{md}=5$\% \hspace{1mm} (SNR of $-3.5$~dB).
\end{itemize}

\begin{figure}[t!]
\centering
\minipage[t]{0.5\linewidth}
\begin{subfigure}[t]{1.0\linewidth}
    \includegraphics[width=\textwidth]{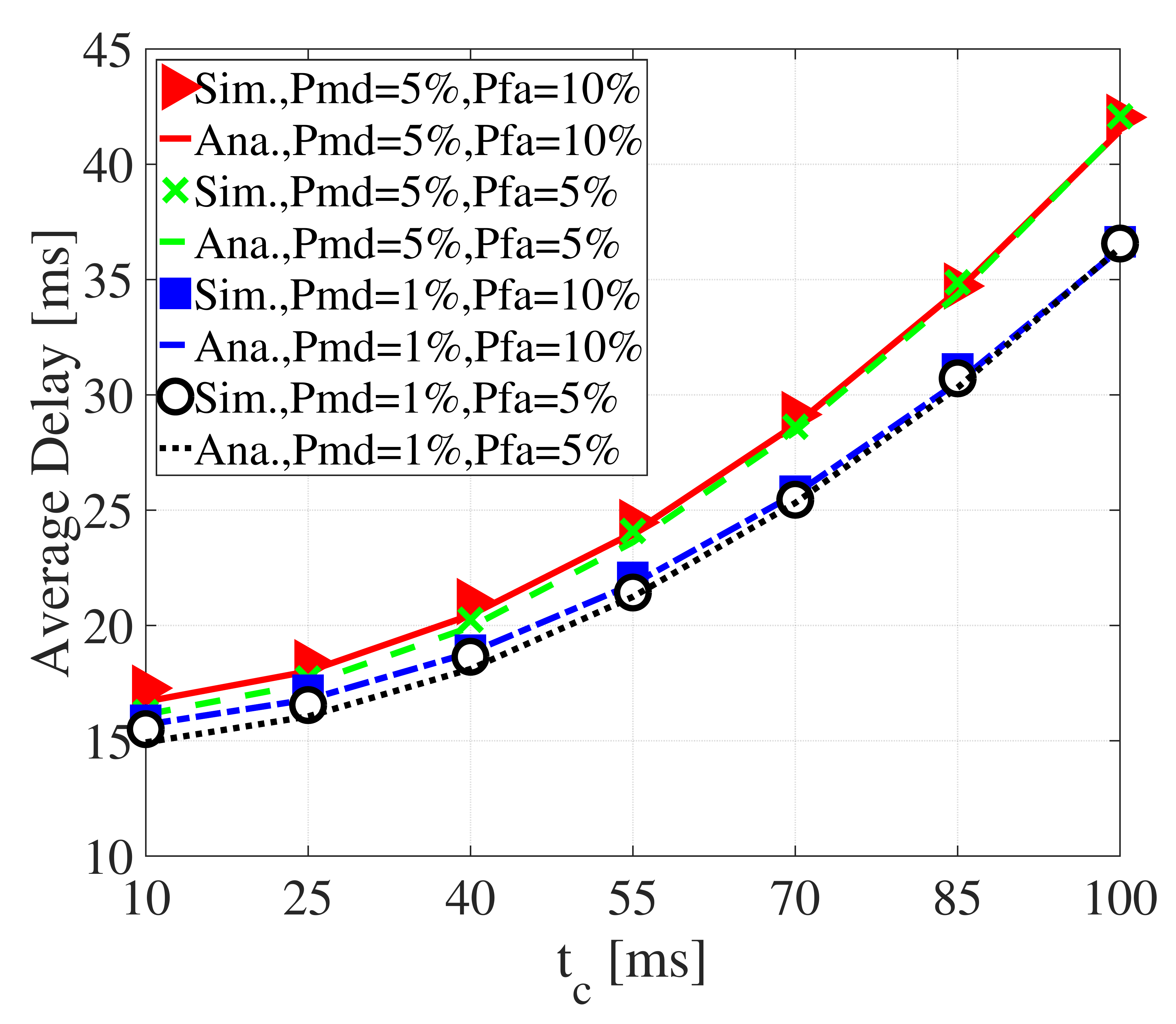}
  \caption{}
\end{subfigure}
\endminipage
\minipage[t]{0.5\linewidth}
\begin{subfigure}[t]{1.0\linewidth}
    \includegraphics[width=\linewidth]{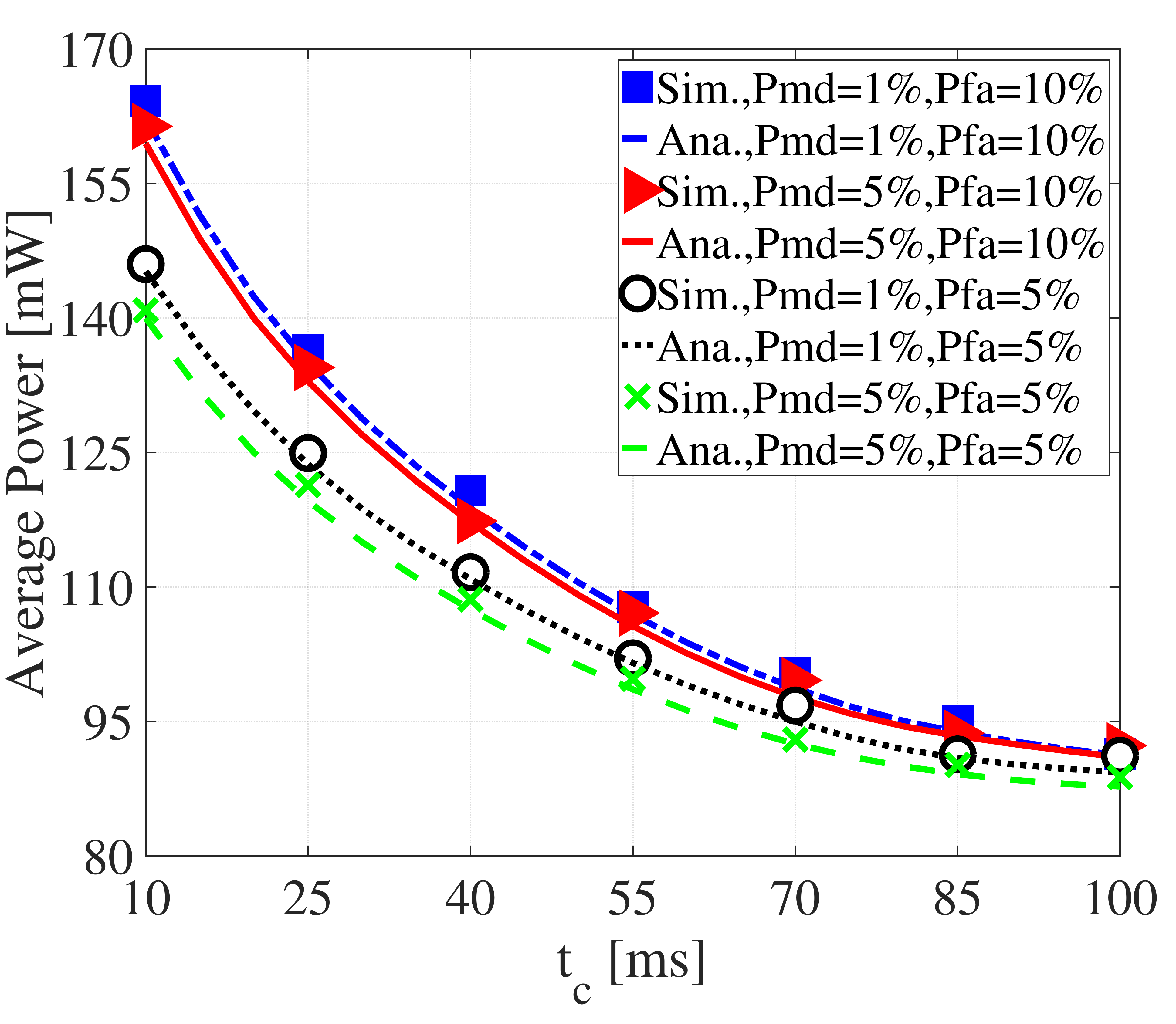}
  \caption{}
\end{subfigure}
\endminipage	
\caption{Analytical and empirical results for the average power consumption and latency characteristics of the  NM for different values of the w-cycle $t_{c}$;  {$T_{ON}=1$ ms and $T_I=12$ ms}.}\label{fig:w_cycle_delay_power}		
\end{figure}

 Fig. \ref{fig:w_cycle_delay_power} then  illustrates the obtained average buffering delay and the  power consumption of the NM for varying values of the w-cycle.  First, it can be observed that the results of the theoretical analysis are very consistent and similar with the empirical ones, for both the average delay and power consumption characteristics, verifying the accuracy of the analytical results. For given values of $T_{ON}$ and $T_{I}$, longer w-cycles increase latency due to longer buffering in each cycle as depicted in Fig. \ref{fig:w_cycle_delay_power}(a), hence shorter w-cycles are favorable for delay sensitive applications. However, reducing the w-cycle increases the energy consumption substantially. The main reasons for such  energy consumption increase in the considered NM concept are  (\emph{i})   more frequent switching of the BBU,  and (\emph{ii}) the  small but nonzero  power overhead   of WRx.  From Fig. \ref{fig:w_cycle_delay_power}, we can also observe that the empirical buffering delay and power consumption values are always slightly higher than the corresponding analytical results. Thus, the analytical models are essentially   lower bounds for the NM delay and power consumption, the main reason for such trend being the simplifying assumptions related to the  packet calls service time made in the analytical work. However, generally speaking, the analytical and empirical results are very well inline.

{\color{black} As it can also be observed in Fig. \ref{fig:w_cycle_delay_power}(a), higher $P_{md}$  for a given $P_{fa}$,  increases   the buffering delay. This is very natural, since misdetection means that the incoming PDSCH data transmission window  is missed. Furthermore, an impact of  $P_{fa}$  on the buffering delay is less tangible. The main reason for this is that after a false alarm, the NM  can still decode   PDWCH. Fig. \ref{fig:w_cycle_delay_power}(b) also clearly shows that a higher $P_{fa}$, for a given $P_{md}$,  increases the energy consumption, due to unnecessarily waking up the BBU.}

\begin{figure}[t!]
\centering
\minipage[t]{0.5\linewidth}
\begin{subfigure}[t]{1.0\linewidth}
    \includegraphics[width=\textwidth]{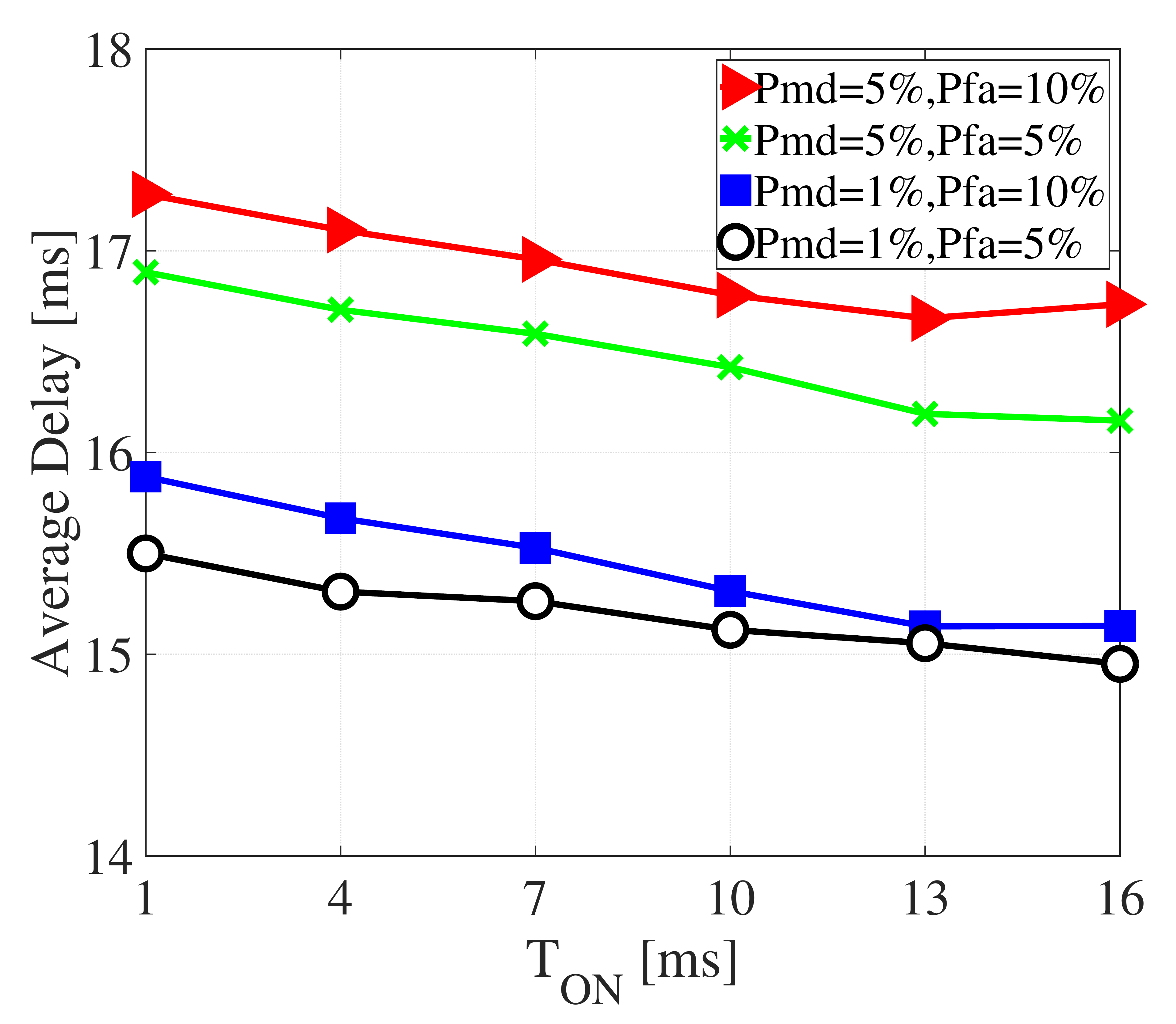}
  \caption{}
\end{subfigure}
\endminipage
\minipage[t]{0.5\linewidth}
\begin{subfigure}[t]{1.0\linewidth}
    \includegraphics[width=\linewidth]{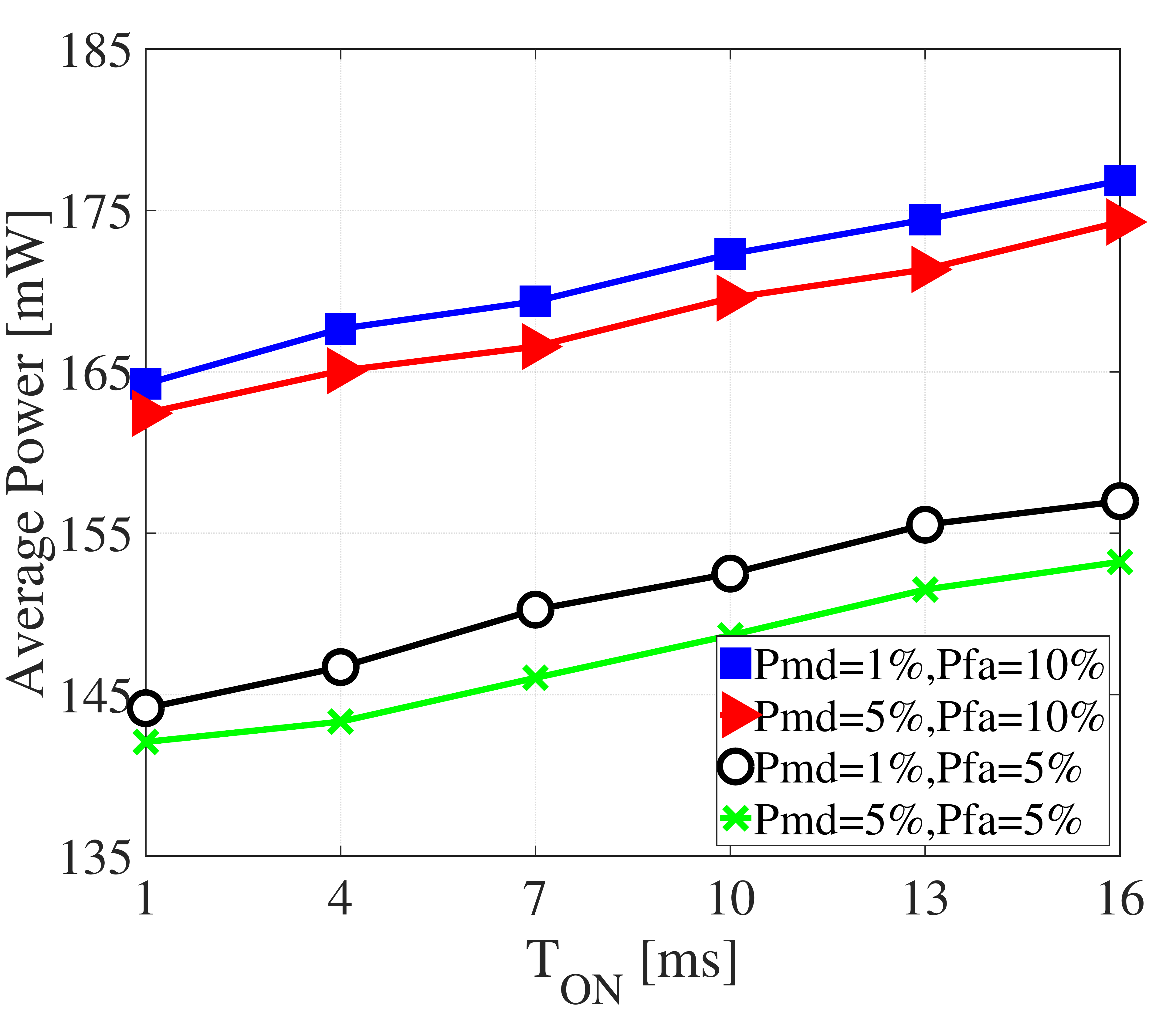}
  \caption{}
\end{subfigure}
\endminipage	
\caption{Empirical results of the average power consumption and latency characteristics of the NM  for the different values of $T_{ON}$;  {$t_c=10$ ms and $T_I=12$ ms}. }\label{fig:t_on_delay_power}			
\end{figure}

 Next, the effects of the on-time duration, $T_{ON}$, on the average delay
and power consumption  are illustrated in Fig. \ref{fig:t_on_delay_power}, for different values of $P_{md}$ and $P_{fa}$. According to Fig. \ref{fig:t_on_delay_power}(a), the value of $T_{ON}$ has a relatively small impact on the average delay. That is, increasing the value of $T_{ON}$ reduces the average delay but only to a small extent. In contrast, as shown in Fig. \ref{fig:t_on_delay_power}(b), the effect of the value of $T_{ON}$ on the power consumption is more notable.  {\color{black} This is because in the event of a false alarm, the UE remains unnecessarily in the active stage for the duration of $T_{ON}$, and thus the larger the value of $T_{ON}$, the larger is the additional energy consumption. Thus, we conclude that overall it is more beneficial to use relatively small values of $T_{ON}$.}

 \begin{figure}[t!]
\centering
\minipage[t]{.5\linewidth}
\begin{subfigure}[t]{1.0\linewidth}
    \includegraphics[width=\textwidth]{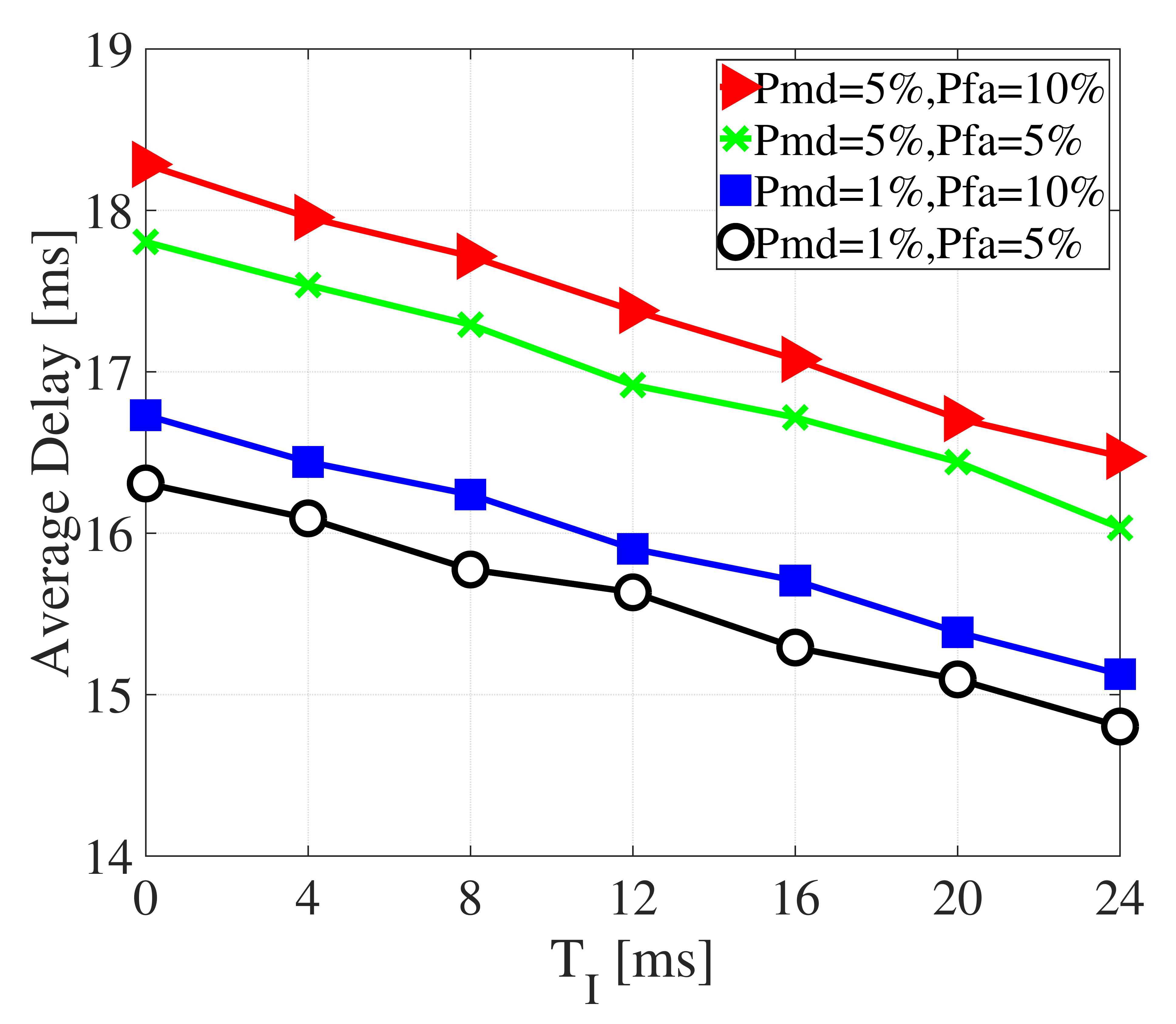}
  \caption{}
\end{subfigure}
\endminipage
\minipage[t]{0.5\linewidth}
\begin{subfigure}[t]{1.0\linewidth}
    \includegraphics[width=\linewidth]{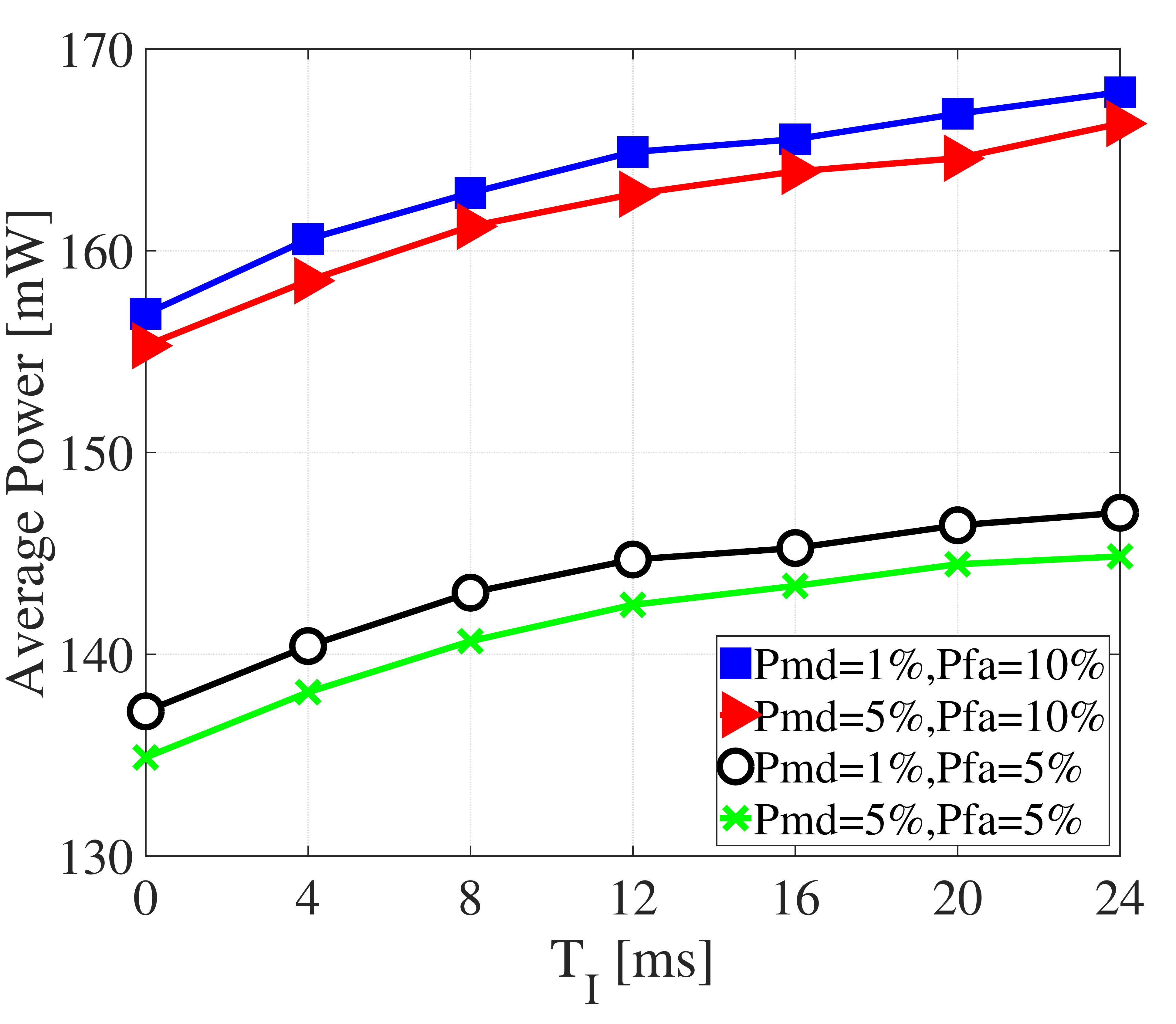}
  \caption{}
\end{subfigure}
\endminipage
\caption{Empirical results of the average power consumption and latency characteristics of the NM  for the different values of $T_{I}$;  {$t_c=10$ ms and $T_{ON}=1$ ms}.	}
\label{fig:t_i_delay_power}		
\end{figure}

 Fig. \ref{fig:t_i_delay_power} then illustrates and assesses how the value of the inactivity timer, $T_I$, affects the average delay and power consumption under different detection scenarios. From Fig. \ref{fig:t_i_delay_power}(a), we can see that   a longer $T_I$ results in a shorter delay.
However, a longer
$T_I$  results also in a higher probability for wasting energy due to absence of any packet, while BBU is fully active.  
 Moreover, for lower $T_I$, the average delay increases noticeably. The figure thus illustrates the impact and importance of correct configuration of $T_I$ on both the delay and power consumption, depending on traffic parameters.

 \subsection{Comparison with DRX}
Finally, the performances of the developed NM and a DRX-enabled reference cellular subsystem are compared, for different values of the w-cycle  and the short/long DRX cycle lengths. We assume that the w-cycle and the short-DRX cycle vary  {\color{black} from $5$   to $250$ ms and from $15$   to $340$ ms,} respectively. In addition, for fair comparison, we assume that the DRX long cycle is four times of its short DRX cycle, and also that the short DRX timer is $16$~ms. Similar assumptions are made also, e.g., in \cite{Lauridsenthesis}, \cite{DRX_etsi}, \cite{Tseng}.
 
\begin{figure}[t!]
\centering
\includegraphics[scale=.19]{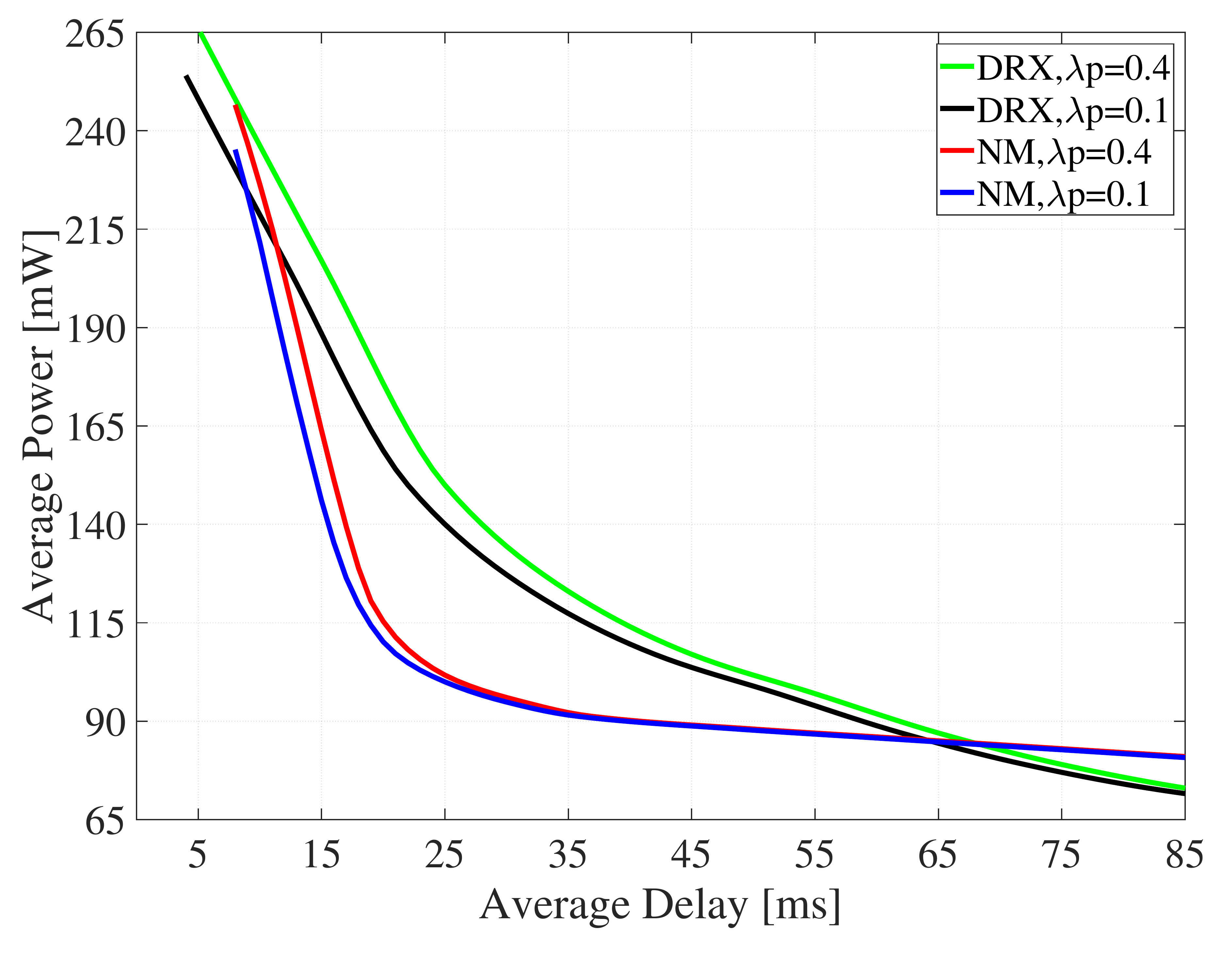}
\caption{Power-delay operating characteristics curve for DRX-UE and NM under different mean packet arrival rates  of $0.1$ and $0.4$  packets per millisecond, assuming $P_{md}=1\%$ and $P_{fa}=10\%$;  {$T_{ON}=1$ ms and $T_{I}=12$ ms}.}
\label{fig:drx_wrx}
\end{figure}

 {\color{black} Fig. \ref{fig:drx_wrx} shows  the achievable average power consumption versus the average delay, for two different values of the mean packet arrival rate. We can observe that when the  w-cycle and the short-DRX cycle are increased, from $5$ to $250$~ms and from $15$ to $340$ ms, respectively, the average packet delay increases while the power consumption reduces. This is natural because when increasing the w-cycle or the short-DRX cycle, the sleep ratio increases, and thus power consumption is reduced but at the cost of an extra delay. Moreover,  for given delay requirement, and per power saving mechanism (wake-up scheme or DRX), power consumption for the higher packet arrival rate ($\lambda_{p}=0.4$ packets/ms) is larger than that for the lower packet arrival rate ($\lambda_{p}=0.1$ packets/ms). This is mainly because the larger value of $\lambda_{p}$  implies more packets arriving at gNB during a given time interval. This, in turn, makes the UE BBU more likely to stay in the active state, leading to a higher power consumption for a given delay constraint.   Interestingly,  for given power consumption, and per power saving mechanism, the delay for $\lambda_{p}=0.4$ packets/ms is larger than that for $\lambda_{p}=0.1$ packets/ms. At first glance, it may sound contradictory to the aforementioned reasoning that UE has lower sleep ratio for higher packet arrival rate. However,  one should keep in mind that in order to have equal power consumption values for two different packet arrival rates,  the   w-cycle or DRX-cycle need to be lower for lower packet arrival rate, and as result the overall delay is shorter for lower packet arrival rate under a fixed power consumption value.}

 In general, as it can be observed in Fig. \ref{fig:drx_wrx}, for delay requirements within 10~ms to 60~ms, the NM consumes much less power than a DRX-based UE. For instance, if the buffering delay constraint is $25$ ms, by utilizing DRX, mobile device may achieve average power consumption of $140$~mW, while with employing NM, this can be reduced down to $100$~mW  at $\lambda_{p}=0.1$. This represents power savings in the order of 30\%, which is a substantial number. It is also observed that  DRX works better for applications which can tolerate very long delays (large sleep-ratio), the reason being that the NM consumes more energy in the  sleep state than DRX based UE in the deep sleep state of long DRX cycle. Though this difference may change if the WRx power consumption is pushed further down. 
 
 {\color{black} Finally, based on Fig. \ref{fig:drx_wrx}, we also acknowledge that DRX  can better satisfy very short delay requirements, mostly due to the fact that the start-up time of cellular subsystem for short-DRX cycle is less than $1$ ms  \cite{Lauridsenthesis}, \cite{DRX_etsi}, \cite{Tseng}.}        {\color{black} In case of the NM, due to the wake-up scheme's offset time, achieving such very low latencies in the order of 1-5~ms is difficult.  Additionally one can observe that for tighter delay constraints,  both methods   consume  higher  amounts of energy, the main reason being that the cellular subsystem  remains in active state more often than in case with longer delay constraints. }

\section{Conclusions}
\label{section:conclusions}

In this article,   a timely wake-up radio based access scheme was studied, developed and analyzed, in order to facilitate improved mobile device power-efficiency vs. latency trade-offs in future 5G and beyond systems. Specifically, an efficient wake-up signaling structure was developed, such that reliable wake-up detection can be achieved even in challenging conditions with negative SNRs. Additionally, efficient receiver processing solutions, building on the proposed wake-up signal structure, were also described, for joint synchronization and wake-up symbol location estimation purposes, as well as to decode the actual wake-up indicator bit. We also addressed the wake-up receiver hardware and implementation aspects, while also analyzed the system power consumption and latency characteristics analytically in closed-form. A large collection of representative numerical results was also provided, evidencing that very reliable wake-up signal detection can be achieved even at negative SNRs. Additionally, the numerical results showed that the delay and power consumption characteristics are functions of the involved wake-up cycle and the related parameters, allowing for the system optimization per application such that desired balance between the power-efficiency and latency is achieved. Finally, it was shown that up to $30\%$ reduction in the mobile device power consumption can be obtained, for given latency characteristics, compared to the existing DRX-based solutions, when latencies in the order of 10-60~ms are considered. {\color{black} Our future work will focus on the wake-up system and WRx parameter configuration optimization for different applications and corresponding traffic characteristics.} {Additionally, further optimized solutions and hardware development are pursued to push down the WRx energy consumption even further.} 

\ifCLASSOPTIONcaptionsoff
  \newpage
\fi
\bibliographystyle{IEEEtran}
\bibliography{IEEEabrv,Master}

 \begin{IEEEbiography}[{\includegraphics[width=1in,height=1.25in,clip,keepaspectratio]{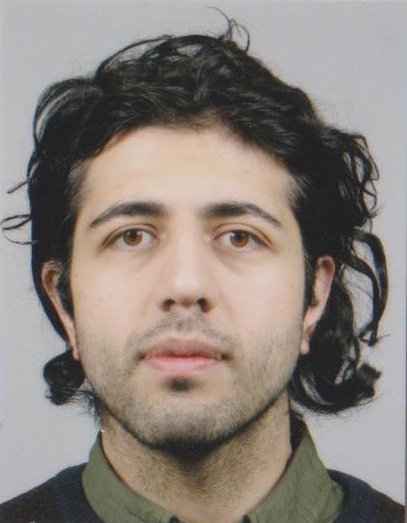}}]
 {Soheil Rostami}   received the M.Sc.   degree in Mobile Communication Systems  with distinction  from the University of Surrey, UK, in 2011. During 2012--2017, he was with the Faculty of Engineering and Science, University of Greenwich, UK, and with Nokia Bell Labs, Belgium. Since 2017, he has been  working as a   Researcher  with 5G  Radio  Network  Technologies  team in Huawei Technologies Oy (Finland) Co., Ltd. In 2018, he was a Visiting   Researcher with   Centre Tecnol\`ogic de Telecomunicacions de Catalunya (CTTC),  Spain. 
\end{IEEEbiography}
\begin{IEEEbiography}[{\includegraphics[width=1in,height=1.25in,clip,keepaspectratio]{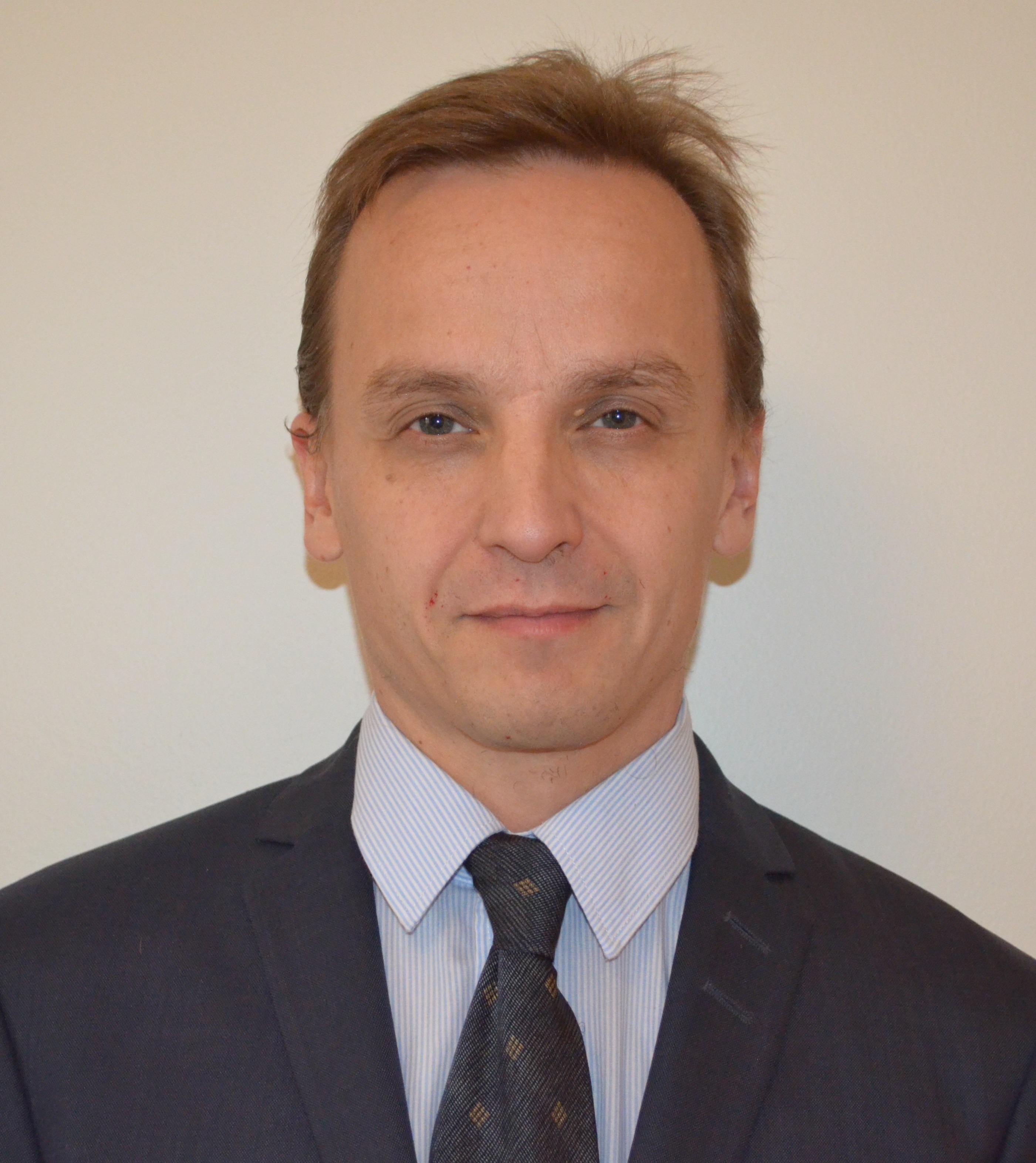}}]
 {Kari Heiska}   received the M.Sc. and Ph.D. degrees from the Helsinki University of Technology (HUT), Finland in 1992 and 2004, respectively. He was with the HUT Laboratory of Space Technology, Nokia Networks, Nokia Research Center and Digita in Finland. Currently he is working as a Senior Researcher at Huawei Technologies in Finland. 
\end{IEEEbiography}
\begin{IEEEbiography}[{\includegraphics[width=1in,height=1.25in,clip,keepaspectratio]{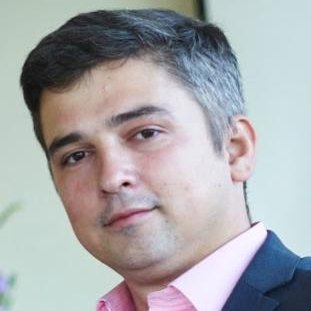}}]
 {Oleksandr Puchko}received the M.Sc degree in wireless communication from Kharkiv National University of Radio Electronics in 2007, and the Ph.D degree in 2013 from the University of Jyvaskyla. During 2008--2017, he was working as a researcher for University of Jyvaskyla,   Magister Solutions, Unikie and Huawei Technologies Oy (Finland). Since 2018 he is Engineer of the Hard Real Time Software for the 5G base station in Mobile Networks at Nokia. 
\end{IEEEbiography}

\begin{IEEEbiography}[{\includegraphics[width=1in,height=1.25in,clip,keepaspectratio]{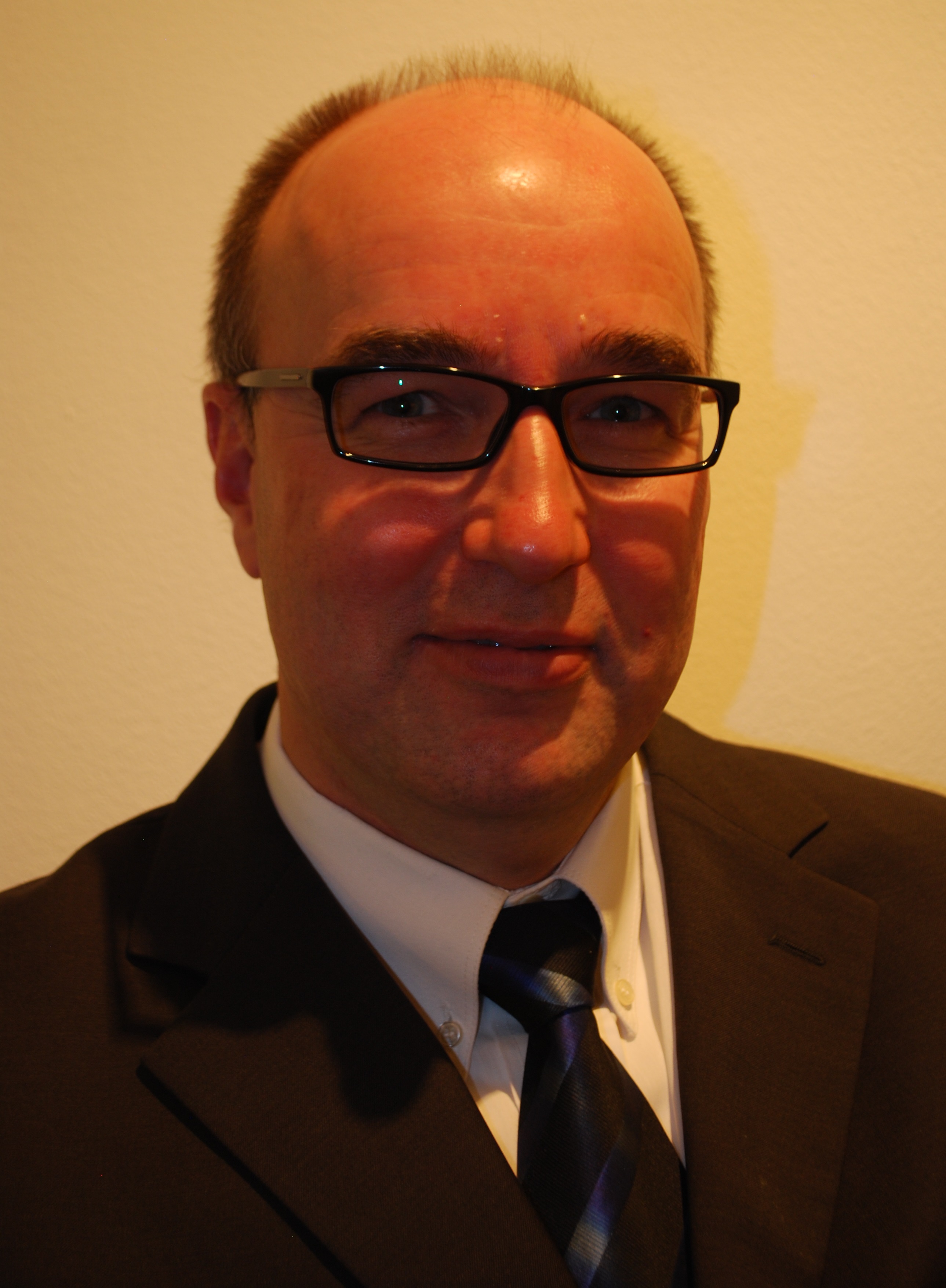}}]%
 {Kari Leppanen} received his M.Sc. and Ph.D. from Helsinki University of Technology, Finland, in 1992 and 1995, respectively, majoring in Space Technology and Radio Engineering. After graduation, he has worked in National Radio Astronomy Observatory (USA), Helsinki University of Technology (Finland), Joint Institute for VLBI in Europe (the Netherlands) and Nokia Research Center (Finland). Currently Kari leads the 5G Radio Network Technologies team at Huawei in Stockholm and Helsinki.
\end{IEEEbiography}

\begin{IEEEbiography}[{\includegraphics[width=1in,height=1.25in,clip,keepaspectratio]{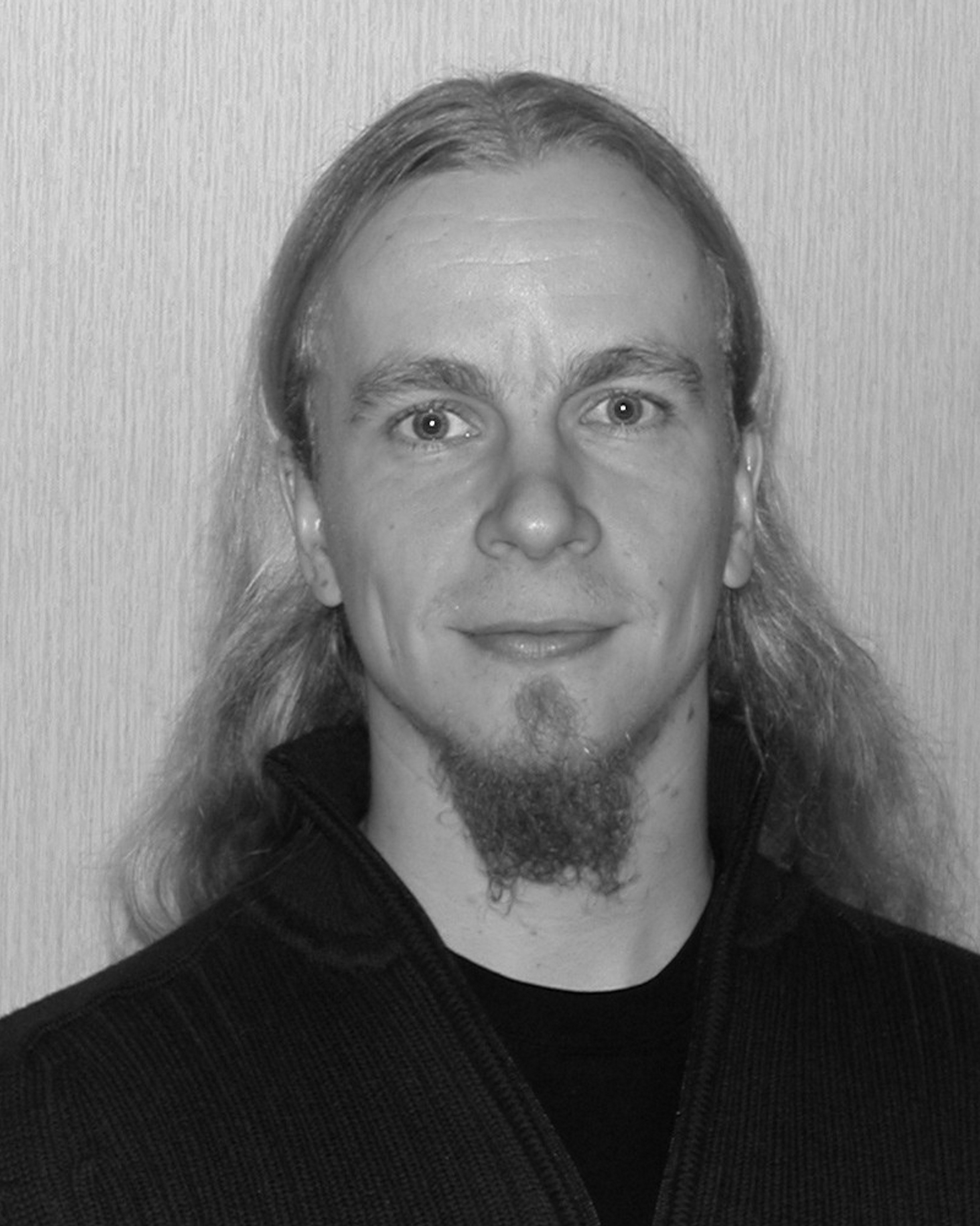}}]%
 {Mikko Valkama}  (S'00--M'01--SM '15) received the M.Sc. and Ph.D. Degrees (both with honors) in electrical engineering (EE) from Tampere University of Technology (TUT), Finland, in 2000 and 2001, respectively.  
 In 2003, he was working as a visiting post-doc research fellow with the Communications Systems and Signal Processing Institute at SDSU, San Diego, CA. Currently, he is a Full Professor and Department Head at the newly established Tampere University (TAU), Finland. His general research interests include radio communications, radio localization and radio-based sensing, with emphasis on 5G and beyond mobile radio networks.
\end{IEEEbiography}

\clearpage

{\color{black}

\section*{Appendix: Further Implementation Details}
This supplementary material provides further implementation details, related to the prototype hardware hardware.

\subsection*{Task Specific Aspects}
The initial synchronization is performed by cross-correlating the signal with a delayed version of itself, and then applying a moving average filter with averaging length equal to the CP length. The maximum correlation magnitude, and the corresponding phase angle, are then measured in every OFDM symbol length. This means that the phase angle from the strongest correlation peak is registered, and is utilized to estimate the frequency offset. In this work,   the hardware architecture proposed in  \cite{Beek1997} is adopted. It requires ca. $3N_{b}(N_{cp}-1)x$ and $N_{b}(3N_{cp}+1)x$ complex additions and  multiplications, respectively.

\thispagestyle{empty} 

To compensate for the CFO, an NCO is utilized to generate a complex phasor at the estimated frequency. The complex conjugated phasor is   multiplied by the received signal to correct the frequency offset. In this work,   the multipartite table method (MTM) is applied    to implement arctangent function and sine/cosine generator. It is proven to be much faster, more accurate and more efficient in power and area compared with some alternative   algorithms \cite{Rust}. The NCO consists of a register accumulating the $16$ bit input values and a phase-to-sine mapper. In order to enable hardware-efficient signal processing, the full input phase range is scaled to $2^{16}$ values, and saved in internal memory of the RFIC. The phase accumulator produces a digital sweep with a slope proportional to the input phase. On the other hand, de-rotation must be performed, which can be easily achieved by vector rotation. For our work, we have selected modified radix-$4$ tree-multiplier structure \cite{Keating}, and segmentation is used.

The computational complexity of the WRx mainly comes from the  IFFT and FFT blocks. In this work,  radix-$2$ single-path delay feedback is adopted due to their ability to use the registers more efficiently by storing the one butterfly output in feedback shift registers. A single data stream goes through the multiplier at every stage \cite{Shousheng}. This algorithm requires $\log_2 (N-2)$    multipliers, $\log_2 N$ radix-$2$ butterflies and  $N - 1$ registers. Its memory requirements are minimal.

Specifically, as detailed in Section \ref{section:system}, $a$ frequency-domain correlators with size of $K$   are needed to detect the parameter set. Correlations can generally be done using multiple IFFTs, at the expense of an increase in hardware complexity and cost, or a single  IFFT at the expense of an increase in decision delay. The latter is appropriate for scenarios where the frequency drift is low, or the value of $a$ is low. Due to fact that ZC generation in real-time is computationally expensive, in our design the  ZC sequence pre-computed by the BBU  is buffered in the available memory of the RFIC to be used in the WRx.

In the WRx implementation, the BB signal is sampled at a rate of $f_s=N \Delta f=1.92 $ MHz,  and converted to the frequency-domain through a $128$-point FFT unit ($N=128$).  The PDWCH uses $K=117$ subcarriers for the signaling of the WI bits, and $N_g=10$ as guard subcarriers.  Due to the use of the FFT and IFFT, the proposed WRx implementation has an  inherent algorithmic latency of $2N=24,576$ clock cycles  at $152$ MHz. Furthermore, its start-up and power-down periods together are less than $0.5$ ms.

\subsection*{Additional Selected Issues}
\noindent\textbf{RFFE:}
An antenna, LNA, and SAW  filter  for channel selection   are the main elements of   RFFE  used in the measurements.

\noindent\textbf{RFU:} The    RFU   downconverts the analog amplified received high-frequency signal to a low-frequency BB signal directly using  single balanced mixer, channel selection filter and AGC; eventually, it passes the I/Q samples to the DFE. In the proposed NM design,  the BB filters for channel selection     are active op-amp based filters known as gyrators. Additionally, the BB    AGC  module monitors and averages the  received signal strength,  and   adjusts the gain of the RFFE.

\noindent\textbf{DFE:} The DFE   consists of ADC, DC offset cancellation, digital filter, and decimator.
 Because of narrow bandwidth  of   wake-up signaling, NM requires very low-lost conversion method; thus Sigma delta converter   is utilized, which provides    high dynamic range and high flexibility in converting diverse bandwidth input signals. Due to zero-IF architecture, DC offset removal is essential. A  finite impulse response (FIR) filter  is used to attenuate signals, and noise that are outside PDWCH subband. Finally,  to eliminate redundant sampled data  at the output of DFE, decimator is employed.

\noindent \textbf{BBU:} The BBU receives the digitized signal as complex I/Q samples from the analog to digital converter  (ADC) via  RF-BB interface. In our design for BB processing,    a set of commercial modules are utilized, as it is seen in Fig.~\ref{fig:block_NM_schematic}.  To support receiver signal processing related tasks, one high-end DSP core (CEVA-XC$4500$) is employed along with a real-time processor (CEVA-X$4$) for the physical layer coordination between the transmit and receiver subsystems, as well as overall BBU control. In conjunction,  function-specific hardware accelerators are applied for different processing stages such as FFT, de-interleaver,  ILR and etc. Additionally, for protocol stack execution, a quad-core high-performance real-time processor (Cortex R$8$) is employed.

\noindent\textbf{Digitally Controlled Crystal Oscillator, DCXO:} NM uses a single high frequency crystal oscillator to generate accurate master clock ($38.4$ MHz).   DCXOs  at the clock interface are utilized to adjust the reference frequency.  Furthermore, frequency offsets are measured by   BBU and WRx, and compensated through  automatic frequency correction (AFC) module based on BBU and WRx.

\noindent\textbf{Frequency Generation Unit, FGU:} Different frequencies required in the NM are derived from the master clock  using multiple   PLLs. Local oscillator (LO) frequency for the mixer,  the sampling frequency at the ADC, the clock signal for  timer/counter,  DFE, WRx and BBU    are the main units that   receive the clock. }
\end{document}